\documentclass[10pt,journal,compsoc]{IEEEtran}
\ifCLASSOPTIONcompsoc
  \usepackage[nocompress]{cite}
\else
  \usepackage{cite}
\fi
\ifCLASSINFOpdf
  \pdfoutput=1\relax                   
  \usepackage{graphicx}                
  \usepackage{xcolor}
  \usepackage{amsmath}
  \usepackage{amssymb}
  \usepackage{tabularx}   
  \usepackage{algorithm}
  \usepackage{algorithmic}
  \usepackage{multirow}
  \DeclareGraphicsExtensions{.pdf,.png,.jpg,.jpeg} 
\else
  \usepackage{graphicx}                
  \DeclareGraphicsExtensions{.eps}     
\fi%

\usepackage{hyperref}

\graphicspath{{figures/}{pictures/}{images/}{./}} 

\usepackage{microtype}                 
\PassOptionsToPackage{warn}{textcomp}  
\usepackage{textcomp}                  
\usepackage{mathptmx}                  
\usepackage{times}                     
\usepackage{cite}                      
\usepackage{tabu}                      
\usepackage{booktabs}                  
\usepackage{amsmath}
\usepackage{amsmath}
\newcommand{\clr}{\textcolor{black}}

\begin{document}
%
\title{Local Latent Representation based on Geometric Convolution for Particle Data Feature Exploration}
%
%
%
%

\author{Haoyu~Li
        and~Han-Wei~Shen
\IEEEcompsocitemizethanks{\IEEEcompsocthanksitem H. Li is with the Department
of Computer Science and Engineering, The Ohio State University, Columbus,
OH, 43210.\protect\\
E-mail: li.8460@osu.edu
\IEEEcompsocthanksitem H. Shen is with the Department
of Computer Science and Engineering, The Ohio State University, Columbus,
OH, 43210.\protect\\
E-mail: hwshen@cse.ohio-state.edu}
\thanks}

\IEEEtitleabstractindextext{%
\begin{abstract}
Feature related particle data analysis plays an important role in many scientific applications such as fluid simulations, cosmology simulations and molecular dynamics. Compared to conventional methods that use hand-crafted feature descriptors, some recent studies focus on transforming the data into a new latent space, where features are easier to be identified, compared and extracted. However, it is challenging to transform particle data into latent representations, since the convolution neural networks used in prior studies require the data presented in regular grids. 
In this paper, we adopt Geometric Convolution, a neural network building block designed for 3D point clouds, to create latent representations for scientific particle data. These latent representations capture both the particle positions and their physical attributes in the local neighborhood so that features can be extracted by clustering in the latent space, and tracked by applying tracking algorithms such as mean-shift. We validate the extracted features and tracking results from our approach using datasets from three applications and show that they are comparable to the methods that define hand-crafted features for each specific dataset.

\end{abstract}

}

\maketitle

\IEEEdisplaynontitleabstractindextext

%
\IEEEpeerreviewmaketitle

\IEEEraisesectionheading{\section{Introduction}\label{sec:introduction}}

%
%
%
%
\IEEEPARstart{W}{ith} the growing size and increasing complexity of simulation outputs, feature extraction and tracking have become essential tasks for scientific data analysis. Through extraction and tracking of salient features, scientists can better understand the evolution of scientific phenomena. 
In the domain of data visualization, visualizing extracted features avoids 3D occlusion and improves visualization efficiency, especially when the dataset is large. 
From the early work\cite{Samtaney1994,Ji2003} to the more recent approaches such as topology analysis\cite{viscous_finger_topology,scicon2016} and deep learning\cite{Monfort2017}, feature extraction and tracking methods generally share a common strategy, which is to find the spatially unique and temporally coherent regions based on some feature specific attributes such as variable values, size, shape, topology or texture.

There have been a plethora of studies that focus on designing and engineering feature descriptors in different datasets from various domains\cite{Xu2019,caban2007texture,jiang2002novel,rockstar,viscous_finger_topology,scicon2016,correa2008size}. However, finding a precise definition for salient features can be difficult due to the complexity of many scientific phenomena. Extracting features with hand-crafted feature descriptors can involve tedious manual intervention and is not always robust.
In order to solve this problem, automatic feature extraction approaches based on machine learning, or more specifically neural networks, have been explored. A new representation, called latent vectors, is produced from the raw data using a neural network. After that, feature exploration is performed based on analyses of clusters and distances between the clusters in the latent space. Automatic feature extraction studies showed promising results with little human intervention \cite{Cheng2019,FlowNet}.

When it comes to particle data, such as those produced by molecular dynamics, cosmology, or fluid flow simulations, applying neural network based automatic feature exploration is not a trivial problem. Most of the previous works use convolution kernels to automatically learn the feature descriptors. This requires a regular grid mesh around each pixel/voxel, which is not present in particle datasets. Although there are workarounds that first re-sample the data into a regular grid and then apply convolution \cite{FlowNet}, re-sampling data may cause loss of details or add a significant storage cost. Most importantly, particle geometry defined by the particle positions is lost in this process. Many features, especially those in cosmology simulations, are related to particle geometries, e.g. the density of the particles or the shapes that particles form.

To tackle the challenges, we propose the use of Geometric Convolution (GeoConv) for producing latent representations that integrate the positional information and physical attributes for particle data.
In the booming research field of neural networks for point clouds, where the main goal is to classify shapes formed by 3D point clouds, GeoConv is the key component used in one of the state-of-the-art network architectures\cite{GeoConv}. Similar to the classical Convolutional Neural Networks (CNN), GeoConv builds convolution kernels to detect features using the local information. However, since the neighborhood of particle data or point clouds is not positioned regularly, GeoConv kernel weights depend on the relative position of the point or particle to the kernel center. 

Similar to previous latent space analysis works for visualization, we use an autoencoder with GeoConv to create a latent representation for particle data, which captures the features from the raw input data without any supervision. 
One particle and its local neighborhood, referred to as a particle patch, is processed by the autoencoder to create a latent vector which is a succinct presentation for the particle patch.
This succinct latent representation can then be used for feature exploration.

We show that clusters in the space of our latent representation for the particle data capture dataset-specific features, and that these clusters separate into distinct features.
We create a visual analytics system to allow users to cluster the latent vectors hierarchically and to identify the features interactively.
Additionally, we show that this latent representation can be utilized for feature tracking by applying the mean-shift algorithm to match latent space distributions across time steps.
We summarize the contribution of this study as follows:

\begin{itemize}
\item A neural network based approach specifically designed for particle data, which transforms particles into a local latent representation where the positional and physical attribute features from the particles are preserved.
 
\item A demonstration that this latent representation is useful for feature extraction and tracking.

\item A visual analytics system that helps users interactively explore the produced latent representations and detect the features of interest.
\end{itemize}

\section{Related Works}
Our work contributes to feature extraction and tracking using latent space methods on particle datasets.
We, therefore, review related works to these fields.

\textbf{Latent Space Approaches in Visualization:} 
The approach proposed in this paper is inspired by the study from Cheng et al.\cite{Cheng2019}, where supervised deep learning neural networks are used to extract meaningful features to help the design of a good transfer function for volume rendering. 
Besides this work, many recent studies in visualization utilize latent space exploration and analysis in their application. 
For volumetric time-varying data, Porter et al. \cite{porter2019deep} produce the latent representation for volumes from different time steps, and select representative time steps using latent vectors. 
For streamlines and stream surfaces, Han et al.\cite{FlowNet} train a CNN based autoencoder with voxelized and down-sampled data. Their approach avoids the problem of defining feature descriptor for lines or surfaces but inevitably introduces error in the process of voxelization and down sampling. Meanwhile, the computation and storage overheads of their method are also large. 
Another related recent study by Han et al. \cite{v2v} uses the proximity in the latent space to select the best pairs of variables to perform volumetric data transformation. 
These latent space related works all use a CNN to produce latent vectors. This requires the dataset to be volumetric or to be re-sampled to a volume. 

Meanwhile, recent works in the deep learning domain have designed and applied neural networks\cite{Qi2017,Qi2017a,DGCNN,unsupervised-pointcloud} to point cloud data and have seen great success for classification and segmentation, where both require meaningful feature extraction from point clouds. 
Applying neural networks designed specifically for point cloud data will not be prone to the error incurred from the re-sampling process as the mentioned approaches do.

\textbf{Feature Extraction and Tracking:}
Feature extraction and tracking are two applications for our particle latent representation. We summarize the previous works on feature extraction and tracking in this section. 
First, there are a group of studies focused on attribute space analysis, for example, feature level-sets\cite{Hotz2020} select traits defined by arbitrary geometries in the attribute space and perform iso-surface extraction for multi-variate data, and Linsen et al.\cite{Linsen2008} used a region growing method in the attribute space to solve the problem of extracting features defined over multiple variables in a particle dataset. These works are similar to ours in the sense that we all extract features in the spatial domain based on the selection in another space, which is either a latent space produced by neural networks or the original attribute space. Attribute space methods are possible to be directly applied to particle datasets. However, compared to the attribute space for particle datasets, a latent space is able to capture more information such as the particles' spatial distribution and the relation between positions and attributes.  

Most of the other feature extraction and tracking techniques that do not focus on attribute space fall into two categories. 
The most prevalent approaches\cite{Samtaney1994,Dutta2016,Xu2019,Ji2006,Silver1998,Chen2003} perform extraction and tracking separately. Features are either extracted by spatially coherent attributes or extracted by domain specific hand-crafted feature descriptors. After that, tracking is done by matching the features using spatial overlap over time. 
Other methods\cite{weber2011feature,Ji2003,Sauer2017,Muelder2009} consider features directly in the spatial-temporal space and extract them through topology analysis or high-dimensional iso-surfacing. 
However, most of these feature exploration methods cannot be directly applied to particle data due to the unstructured nature and the lack of connectivity information among the particles. So there have been some methods designed specifically for particle datasets.
Lukasczyk et al.\cite{viscous_finger_topology} resampled irregular volumetric points into a regular volume before the topology analysis is applied to the data set. Similarly, the 2016 SciVis contest winners\cite{scicon2016} proposed an approach that constructs a smooth scalar field before using a contour tree to extract critical points. 
In the dataset for 2015 SciVis Contest, a dark matter halo merger tree that records how halos evolve is obtained using the method by Takle et al.\cite{takle2012tracking}. \clr{Han et al.\cite{han2021exploratory} perform flow map interpolation using ML model on unstructured particle data.} The work by Dutta and Shen\cite{Dutta2016} utilized an incremental Gaussian Mixture Model to both detect and track features. 
Although their experiments are done with volumetric datasets, their method can be adopted to particle data.

\textbf{Point Cloud Neural Networks and their Visualization Application:}
A deep learning survey for 3D point clouds provides a comprehensive overview of the recent point cloud neural network studies\cite{Guo2019}.
The PointNet\cite{Qi2017a} is one of the first methods in deep learning for point cloud data. Subsequently,  Qi et al. expanded the PointNet with PointNet++ by applying the PointNet recursively on a nested partitioning of the input point cloud to improve efficiency and robustness\cite{Qi2017}. PointNet++ borrows the ideas of hierarchically applying filters to extract both local and global features from CNN.
Other recent works improve on the basic architecture of PointNet++\cite{unsupervised-pointcloud,DGCNN,Thomas2019,GeoConv,PointConv,Li2018,Hermosilla2018}. In our work, GeoConv \cite{GeoConv} is adopted in our neural networks as it shows good performance with a relatively small network.
LassoNet \cite{Chen2020}, which solves the problem of interactive selection of objects with the lasso is one of the few visualization works we can find that utilizes 3D point cloud neural networks.


\section{Learning Local Latent Representation}
\label{latent vector} 
In this section, we describe how we generate latent vectors for particle datasets using an autoencoder. 
A classical autoencoder is composed of two parts: the encoder, whose objective is to compress the original data into a reduced dimensional representation called a latent vector, and the decoder, which takes the latent vector as input and reconstructs the original data.
Scientific data can be prohibitively large to compute on directly. Therefore, we instead identify salient features in small local regions by organizing the particles into local patches.
This not only reduces the memory working set size but also makes it easier for neural networks to learn the local representation of particles. 
A particle patch is defined as follows. Considering a center location $\vec{p}\in\mathbb{R}^{3}$ is given for a 3D particle data, the corresponding particle patch $N(\vec{p},r)$ is defined by all the particles whose distance to $\vec{p}$ is less or equal to a distance $r$: $N(\vec{p},r)=\{\vec{q} \mid \lVert \vec{p}-\vec{q} \rVert \leq r \}$. 

The autoencoder is used to transform a particle patch into a latent vector.
The feature descriptors suitable for the dataset are automatically learned by the encoder, and the latent vector represents the high-level features detected by the learned feature descriptors. Therefore, the latent space has two characteristics that make it more appropriate for feature-related analysis.
First, noise in particle patches has less influence on the latent space compared to the influence on the original particle representation\cite{bengio2013representation}. Based on other training samples seen, the encoder detects the most similar feature by the learned feature detection kernel and ignores the noise. 
This makes the latent vector more stable than raw positions and attributes. This effect can be observed from our reconstruction result in \autoref{sect:recon}.
Second, the distances in latent space are more consistently related to the presence of high-level features compared to the attribute space for the particle patches because they are encoded by these learned high-level feature detectors. This characteristic of latent spaces is also found in other works utilizing latent representations \cite{FlowNet,v2v,porter2019deep}.
Next, we provide the detailed architecture of our autoencoder model.

\subsection{Geometric Convolution and Deconvolution} 

\begin{figure}
 \centering
 \includegraphics[width=\columnwidth]{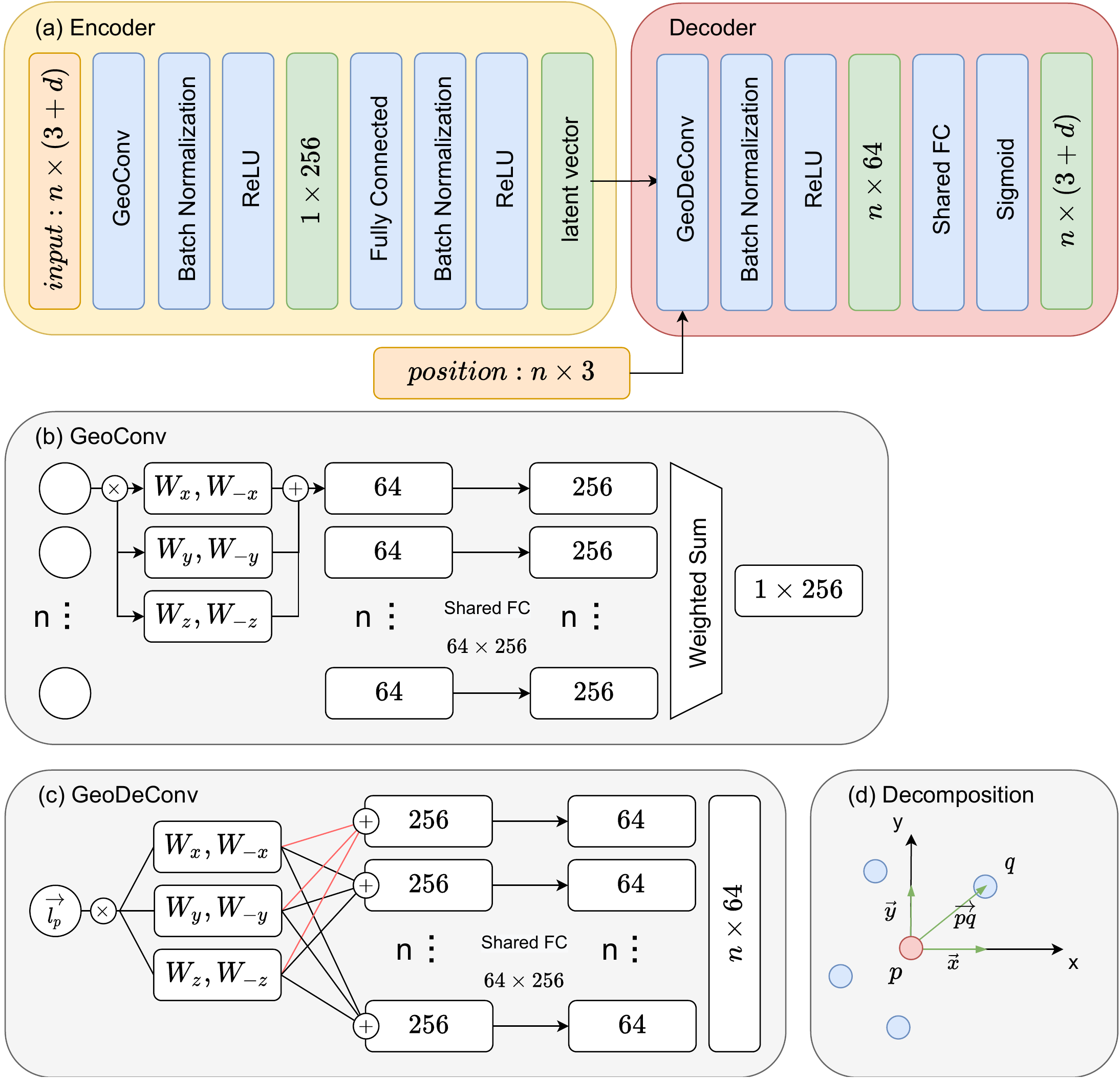}
 \caption{
    (a) The architecture of our autoencoder. The input is a particle patch with $n$ particles, each of which has a 3D position and a $d$-dimensional attribute. 
    (b) GeoConv architecture. The node list on the left stands for the input $n$ particles. The $W$ blocks refer to \autoref{eq1} and the weighted sum block refers to \autoref{eq2}.  
    (c) GeoDeconv architecture. The $W$ blocks refers to \autoref{eq3}. 
    (d) Decomposition blocks show a 2D example of GeoConv, where $\protect\overrightarrow{pq}$ is projected onto two orthogonal bases.
 }
 \label{fig:arch}
\end{figure}

Unlike image or volume data, particle data do not have meshes to connect the particles, hence the convolution kernels used for feature extraction like those in CNNs cannot be easily applied. 
In this work, we adopt an elegant solution called GeoConv\cite{GeoConv} whose core idea is to project the coordinates of a point into orthogonal directions and define a kernel on those directions.
The full architecture of GeoConv is shown in \autoref{fig:arch} (b). Here, we briefly explain the basic building block of GeoConv that is used to construct our autoencoder. 
We can define any particle dataset as a mapping from spatial location to $d$ particle attributes, $f:\vec{p}\mapsto X_{\vec{p}}$, where $X_{\vec{p}}\in\mathbb{R}^d$ is the physical attribute at the particle location $\vec{p}$.
In a particle patch $N(\vec{p},r)=\{\vec{q} \mid \lVert \vec{p}-\vec{q} \rVert \leq r \}$, vector $\overrightarrow{pq}$ is projected into three orthogonal bases $\{ \vec{x}, \vec{y}, \vec{z}\}$, and the projected norm along each basis represents the ``energy" in that direction. We aggregate the attributes of a particle $\vec{q} \in N(\vec{p},r)$ into the center $\vec{p}$ based on its energy in different directions. To differentiate positive and negative directions, we define the following six bases in 3D: $B=\{(0,0,1),(0,0,-1),(0,1,0),(0,-1,0),(1,0,0),(-1,0,0)\}$. Then the aggregation of $q$'s attributes to the center $p$ is as following: 
\begin{equation}
    g(\vec{p},\vec{q}) = \sum_{\vec{b}\in B}{cos^2(\overrightarrow{pq},\vec{b}) W_{\vec{b}} X_{\vec{q}}}
    \label{eq1}
\end{equation} 
where $\vec{b}$ is one of the basis vectors, $W_{\vec{b}}$ is the learnable weight matrix in the corresponding direction and $X_{\vec{q}}$ is the vector of attributes for particle $\vec{q}$. A 2D example of this projection can be found in \autoref{fig:arch} (d). 
The result of $g(\vec{p},\vec{q})$ is a vector, and vectors for each particle in the patch are sent through one shared fully connected layer (shared FC). We use the term shared FC to refer to a fully connected layer whose weight matrix is shared among different particles. As suggested in the original GeoConv study \cite{GeoConv}, the dimensionality of the resulting vector $g(\vec{p},\vec{q})$ and the fully connected layer is set to 64 and 256 . This means the weight matrix in GeoConv shared FC (\autoref{fig:arch} b) is $64 \times 256$ in size, and each of the $n$ particles are multiplied with the same matrix. Shared FC is efficient and can be implemented with $1\times1$ convolution.
Finally, we calculate a vector representation for a particle patch as a summation of all particles in the patch weighted by their distance to the center as shown in \autoref{eq2} and \autoref{eq:kernel}. This last step is illustrated in \autoref{fig:arch} (b) weighted sum block.
\begin{equation}
    GeoConv(\vec{p},r) = \frac{\sum_{\vec{q}\in N(\vec{p},r)}{d(\vec{p}, \vec{q}, r)g(\vec{p}, \vec{q})}}{\sum_{\vec{q}\in N(\vec{p},r)}{d(\vec{p}, \vec{q}, r)}}
    \label{eq2}
\end{equation} 
\begin{equation}
    d(\vec{p}, \vec{q}, r)= \begin{cases}
    (r-\lVert \vec{p}-\vec{q} \rVert)^2, &  \text{for $\lVert\vec{p}-\vec{q}\rVert < r$} \\
    0, & \text{for $\lVert\vec{p}-\vec{q}\rVert > r$} \\
    \end{cases}
    \label{eq:kernel}
\end{equation} 
Here, distance function $d(\vec{p}, \vec{q}, r)$ gives more weights to the particles near the center. 

Similar to the idea of GeoConv, we propose a new building block for particle neural networks called GeoDeConv. GeoDeConv takes as input a latent vector for the center particle of a patch and relative coordinate for every particle in the patch and outputs a new vector for every particle. This vector is then used to reconstruct the positions and attributes for each particle through shared FCs.
We also define the learnable parameters in the fixed orthogonal direction. The GeoDeConv process can be regarded as a dispersion of the latent vector to the neighbor particles. The new vectors $GeoDeConv(\vec{q},\vec{p})$ given the patch center $\vec{p}$ is calculated by: 
\begin{equation}
    GeoDeConv(\vec{q},\vec{p}) = \sum_{\vec{b}\in B}{cos^2(\overrightarrow{qp},\vec{b}) W_{\vec{b}} l_{\vec{p}}}
    \label{eq3}
\end{equation} 
where $l_{\vec{p}}$ represents the input latent vector, $B$ represents the same six orthogonal bases, and $W_{\vec{b}}$ are similarly defined as weight matrices in different directions. However, they are not the identical weights used in the GeoConv. We illustrate GeoDeConv in \autoref{fig:arch} (c).

\subsection{Particle Autoencoder} 

Having the GeoConv and GeoDeConv as building blocks for the neural network, we propose an autoencoder architecture to transform a particle patch into a latent vector. 
Our autoencoder architecture is composed of two parts: the encoder and the decoder.
Our encoder takes as input a particle patch defined by all the particle positions and related physical attributes and produces a single latent vector $l_{\vec{p}}$. The decoder takes $l_{ \vec{p}}$ as input and reconstructs the positions and attributes in the particle patch. We present the architecture of our autoencoder in \autoref{fig:arch} (a).
Since the particle patch size is relatively small in our experiments, we only use one layer of GeoConv followed by one fully connected layer, and the decoder is constructed using one layer of GeoDeconv followed by one shared FC. 
Batch normalization is applied after each layer to stabilize the training and to converge faster\cite{bn}. We use ReLU as the activation function except for the last layer\cite{relu}. Sigmoid is used at the last layer to constrain the output value between $[0,1]$.
To train our network, we use a mean squared error loss, measured between the ground truth attributes and the reconstructed attributes. In our experiment, we choose training batches to be as large as our memory constraint allows.

\subsection{Data Sampling and Preprocessing}
\label{traning_testing}
\begin{figure} 
 \centering 
 \includegraphics[width=\columnwidth]{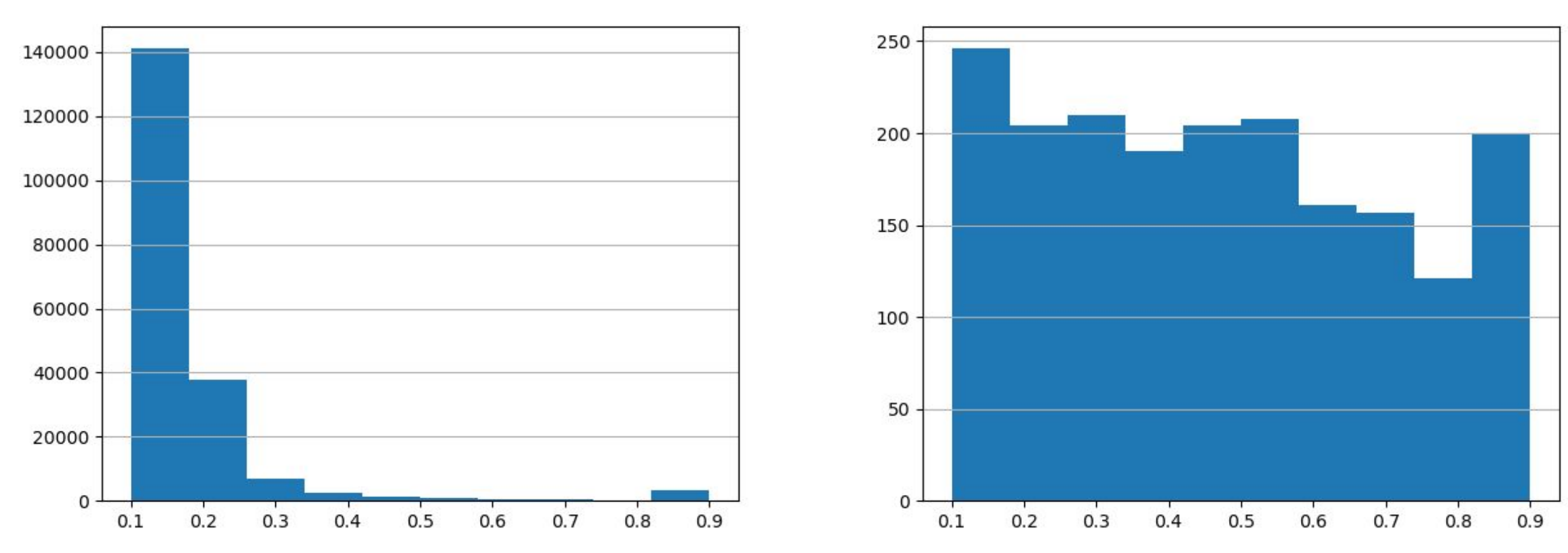}
 \caption{Left: Distribution of salt concentration of all particles (about 190k particles in total). Right: Distribution of the sampled particles (1900 samples) }
 \label{fig:sample_strategy}
\end{figure}

To improve network training, we train our network with samples \clr{from all time steps} of the particle dataset that capture a more diverse feature set than random sampling. We sample based on the distributions of the particle attributes by drawing fewer samples with values that occur frequently, and more samples that have less frequent attribute values.
This sampling rule is based on the observation that very frequent attribute values usually correspond to the background or homogeneous regions with nearly no features. In addition, there is no benefit to train the neural networks with redundant training data. This strategy is also suggested by a recent work of probabilistic data-driven sampling \cite{Biswas2020}.
To realize this sampling strategy, we used a method similar to the value-based sample algorithm in \cite{Biswas2020}. It can be briefly described by first building a histogram in the particle attribute space and then sampling using the inverse density of the histogram as the weight. The attribute distribution from the Salt Dissolution dataset is shown in \autoref{fig:sample_strategy} to demonstrate the effect of our sample strategy. 

There is a computation overhead to find all particles within a distance $r$ from a particle. To locate the particles in a radius efficiently, we perform a spatial partitioning of the particles using a kd-tree\cite{bentley1975multidimensional} before neighborhood query. We normalize all the particle coordinates to be in the range ${[0,1]}$ to avoid having to select radius $r$ specific to the data spatial range.
After the particles in a patch are queried using the kd-tree, since we are only interested in the particles' local relative positions to the center but not their absolute positions, particles in patch $N(\vec{p},r)$ are translated relative to the center particle $\vec{p}$, and then again normalized to be in the range ${[0,1]}$ to match the network output range.

\subsection{Determine the Network Hyper-parameters}
\label{hyper_parameter_method}
There are several hyper-parameters to decide when constructing our network, such as the dimensionality of the latent space and the radius of particle patches. We discuss our approach to determine the parameters here and evaluate the chosen parameters in \autoref{hyper_result}.

We choose the latent dimensionality using a random search strategy. Random samples are drawn from a plausible range of the latent dimensionality (from zero to the dimensionality of input particle patch). The computation budget for each sample is fixed and the parameter that gives the best reconstruction quality is used.
The random search strategy is shown to be more efficient than a grid search of hyper-parameters\cite{bergstra2012random}, can be performed in parallel, and is usually used in neural network designs\cite{Cheng2019}.

In the GeoConv architecture, the radius of the particle patch equals the bandwidth of the GeoConv kernel (defined by \autoref{eq:kernel}).
The size of particle patches is of significance in our approach, since choosing an inappropriate radius $r$ will either overly smooth or add noise to the extracted features. 
We follow the assumption that if the bandwidth is appropriate for estimating a regression function $f:\vec{p}\mapsto X_{\vec{p}}$ between the particle position and its attributes, it will also be suitable to be used in neural networks\cite{warner1996understanding}. We reformulate the problem as a kernel regression estimation. 
The Nadaraya–Watson estimator \cite{nadaraya1964estimating} using our kernel is written as:
\begin{equation}
    \hat{m}(\vec{p};r) = \frac{\sum_{\vec{q}\in N(\vec{p},r)}{d(\vec{p}, \vec{q}, r)X_{\vec{q}}}}{\sum_{\vec{q}\in N(\vec{p},r)}{d(\vec{p}, \vec{q}, r)}},
    \label{eq4}
\end{equation}
where $\vec{p}$ is a point in the 3D spatial domain and $r$ is the kernel bandwidth to be estimated using cross validation. The Least Squares Cross Validation for our Nadaraya-Watson estimator is written as: 
\begin{equation}
    LSCV(r) = \frac{1}{n}\sum_{i=1}^{n}(X_{\vec{p_i}}-\hat{m}_{-\vec{p_i}}(\vec{p};r))^2,
    \label{eq5}
\end{equation}
where $\hat{m}_{-\vec{p_i}}(\vec{p};r)^2$ indicates the leave-one-out estimator without the sample point $\vec{p_i}$.
Instead of using all particles across every time step from the dataset, we sample 1\% of particles from every time step to perform cross validation. The sampling technique is the same as the one presented in \autoref{traning_testing}. We find the particle patch defined by bandwidth $r$ around the sample $\vec{p_i}$ to build the leave-one-out estimator $\hat{m}_{-\vec{p_i}}(\vec{p};r)$ using \autoref{eq4}. Following \autoref{eq5}, we calculate the $LSCV(r)$ for every time step and average the $LSCV(r)$ across time steps to get the final $LSCV(r)$ for the dataset.
Then the golden-section search method\cite{kiefer1953sequential} is applied to search for the optimal bandwidth $r$ which minimizes the LSCV. Because the particle positions are normalized to be in the range ${[0,1]}$, we search the optimal $r$ in this range. The estimated bandwidth is then used for the kernel in GeoConv and as the radius for particle patches.

\section{Feature Extraction and Tracking} 
\begin{figure} 
 \centering 
 \includegraphics[width=\columnwidth]{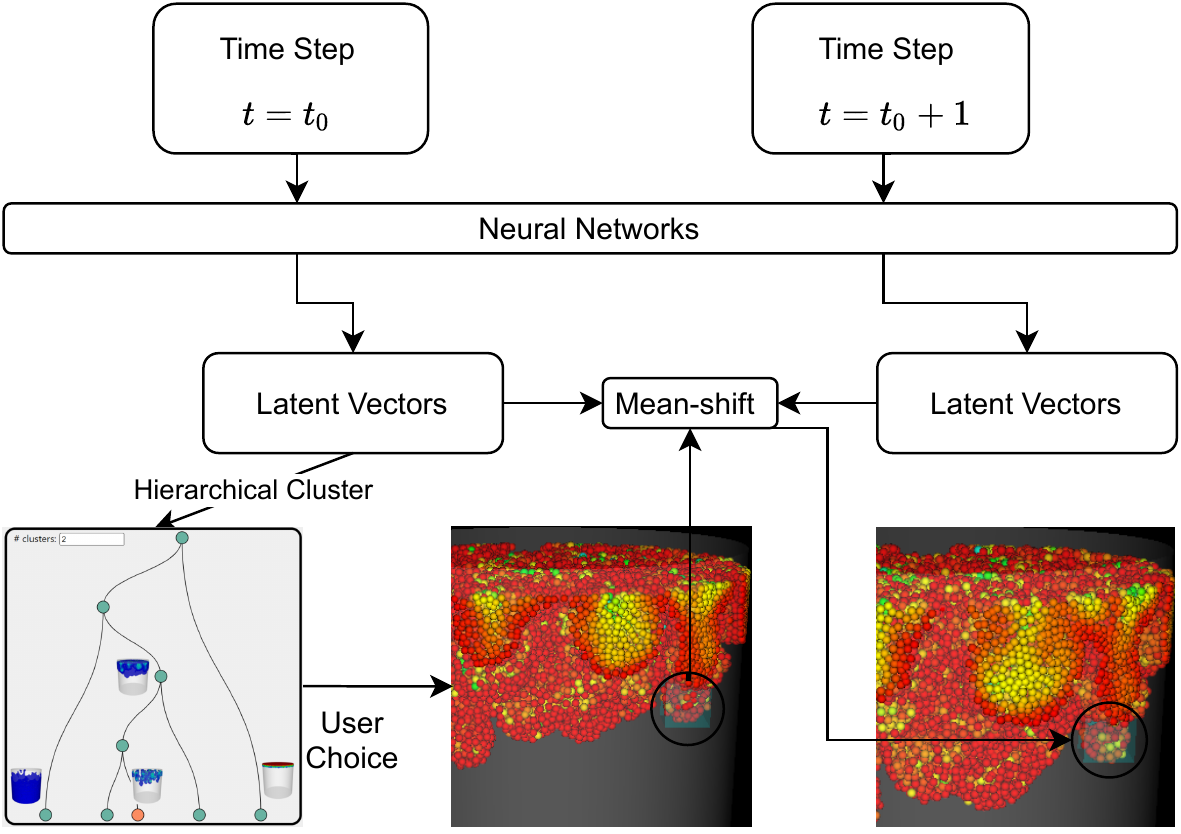}
 \caption{Scheme showing our proposed approach. The neural network is pre-trained with particle patch samples. The tracking we show in the scheme is only for one time step. Actual feature tracking can be repeated until the desired time step is reached. }
 \label{fig:process_overview}
\end{figure}

In this section we introduce our feature extraction and tracking methods in detail.
A workflow of the extraction and tracking process is described in \autoref{fig:process_overview}. 
After the neural network is trained, we can use it to infer the latent vectors for arbitrary particle patches at any selected time step in the domain. As discussed, latent vectors produced by the autoencoder capture high-level features in particle patches. Therefore, patches with similar features will be in close proximity in the latent space. This enables us to explore and find features by analyzing the clusters in this space.
A time step of interest is first selected by users. With the help of our visual analytics system, users identify one or several feature clusters of interest. After that, these clusters can be visualized in the physical space, and users can select any specific region that contains the feature of interest to track over time.  
Features are tracked using the mean-shift algorithm to find the matched location in the consecutive time steps within the desired time span.

\subsection{Latent Space Visual Exploration} 

A typical workflow of feature exploration starts with selecting a time step of interest by the user. This can either be done based on prior knowledge about when the interesting features may occur, or by simply selecting once every few time steps to explore. After the time step of interest is found, our visual analytics system can help the user interactively explore the dataset and find the features of interest \footnote{Source code is available at https://github.com/harviu/particleNN}. This system is composed of three linked parts: a hierarchical cluster view, a 2D feature space view of the latent vectors, and an original physical space view. 
Domain knowledge from the users are injected by modifying the hierarchical clustering, while
feature space and physical space views are updated interactively to provide real-time feedback based on features automatically captured by the latent representation. 


\subsubsection{Hierarchical Cluster View}
Similar to the method used by He et al. \cite{He2020} and Cheng et al.\cite{Cheng2019}, we arrange the latent vectors in a hierarchical clustering tree. The root of this tree represents all the latent vectors. 
Links between parent and child nodes depict splitting the latent vectors from the parent node into multiple clusters using the k-means algorithm.
Whether to split a node and the number of children ($k$ in k-means) to which the parent node splits are decided by the user.
By default, the number of children of each node is set to two but can be increased by the user using the input box based on how many particle groups are shown in the feature space projection (\autoref{sect:feature_space}) of the parent node or when the physical space views (\autoref{sect:physical_space}) of the child clusters are not desirable. We choose the k-means clustering algorithm instead of spectral clustering used by Cheng et al. \cite{Cheng2019} and agglomerative hierarchical clustering used by He et al. \cite{He2020} mainly because of its efficiency on large data.

The hierarchical cluster view provides a vertical tree layout showing the current hierarchy of the clusters. Initially, it's merely a root node representing all particles in the same cluster. After the users choosing to split any node, the view will be updated interactively. The tree layout we choose arranges all the leaf nodes on the same level, which helps the user focus on the clusters of interest they have just created. An example of the hierarchical cluster view is shown in the left half of \autoref{fig:fpm_vis}.

\subsubsection{Feature Space View} 
\label{sect:feature_space}
We adopt t-distributed stochastic neighbor embedding (t-SNE) as our dimensionality reduction technique and visualize the projected latent vectors as 2D scattered points. This visualization only provides a view of the relations between the clusters, and we note that it may not be sufficient for users to decide the features of interest solely by the 2D projection. However, when the 3D visualization in the original physical space is too crowded, 2D feature space projection may reveal the number of possible distinct features. This helps the user decide whether to split a node and how many child nodes it should split into. To avoid applying t-SNE on all the particles which may require extensive computation time, we applied the same sampling technique as used in training data sampling discussed in \autoref{traning_testing} to pick around 1\% of all particles from the selected time step for visualization.

When a node is selected in the hierarchical cluster view, the t-SNE projection of the corresponding particles will be highlighted to help the user focus on the cluster of interest currently. The right panel in \autoref{fig:fpm_vis} presents an example for the feature space view.

\subsubsection{Physical Space View}
\label{sect:physical_space}
This view shows a subset of particles (defined by a selected node) in the original 3D physical space. We assign the colors to the particles which encode a specific scalar attribute of interest. We normalized the coordinates of the particles in the subset to be in the range $[0,1]$, so that the camera can be set to be always facing the center of the visualized particles. 
When users interact with a node in the hierarchical cluster view, the corresponding particles are visualized in this view. Users can refer to the visualization to decide whether to split a cluster into smaller clusters or not and finally spot the desired cluster containing the features of interest. Since the clustering result is influenced by $k$ that users choose, users can refer to the physical space view to validate the parameter $k$ they set when splitting the parent node. If the clustering result is not desirable, the users can easily revoke the splitting operation and test with other more suitable values for $k$. The physical space view is shown in a separate pop-up window in our system. For a better presentation, we display the physical space views beside the cluster nodes and feature groups in \autoref{fig:fpm_vis}.

\subsection{Mean-shift Tracking Using Latent Distribution}

Traditionally, people perform feature tracking by first extracting features in every time step and then matching them across the time steps. In that case, the tracking efficiency is bounded by the extraction efficiency and can be expensive. The situation becomes worse when the latent inference is time-consuming. Therefore, we adopt mean-shift tracking that finds the local region with the highest latent similarity between consecutive time steps.

First, a region of interest is selected by the user with the help of the system mentioned in the previous section. The $v$ dimensional latent vector for every particle in the cubic region is collected. We reduce the latent dimensionality to four using the principal component analysis (PCA) to reduce the computation time for the multi-dimensional histogram and to increase tracking stability. Treating each latent vector as a sample, our feature of interest selected by the user can be represented by a four-dimensional histogram. Therefore, our tracking goal between two consecutive time steps is to identify the cubic region that has the most similar latent histogram in the latter time step. 

The high-level idea of mean-shift tracking introduced in \cite{Comaniciu2000} is that first weights are given to particles in the region of interest according to the comparison between their latent vector with target latent distribution. Then the region is shifted to the higher similarity direction. This is repeated until the region does not move. 

Initially, we assume the spatial location of the feature is the same as that of the previous time step, which is used as the starting point for tracking. 
Then we move the cubic region to the position with most similar distribution using the mean-shift algorithm and follow the parameters suggested in \cite{asvadi20163d}. Repeating the search step on the time sequences will give us the movement of the interested feature through time steps.

\section{Experiments}

We evaluated our neural networks based latent representations using three datasets: Salt Dissolution, Dark Sky, and SNSPH. Before the evaluation, we present the dataset information and model training details in \autoref{sect:dataset}.
Several aspects of our approach are evaluated.
(1) We validate the methods used for determining the hyper-parameters. 
(2) We demonstrate the result of feature exploration in three datasets with the help of a hierarchical cluster exploration tool. The extracted features with our latent representations were verified by comparing with features specifically designed for the dataset or provided by the domain scientists.
(3) We compare the latent representation with alternative methods, local neighbor statistics, and principal component analysis.
(4) We demonstrate that feature tracking using mean-shift on latent space distributions is efficient and accurate.
(5) We detail the computation time and memory use for the proposed approach.

\subsection{Dataset and Experiment Setting}
\label{sect:dataset}
Our approach was evaluated on three datasets generated from the Salt Dissolution simulation (SciVis Contest 2016), the DarkSky simulation  (SciVis Contest 2015), and another supernova simulation called SNSPH. Below we give a brief introduction to the three datasets and their feature extraction target. 

\textbf{The Salt Dissolution} dataset is from the 2016 Scientific Visualization Contest \cite{salt_data}, which models a solid body of salt dissolved by the water using the Finite Points Method. The original data come with 3 different resolutions and 120 time steps each. We tested on the lowest and highest resolution data which have around $190,000$ and $1,800,000$ particles at every time step respectively. Every particle has four-dimensional physical attributes: a salt concentration value and 3D velocity. The feature extraction target is to find features called \textit{viscous fingers}, which is defined only by the salt concentration.

\textbf{DarkSky} is a cosmology simulation dataset from 2015 SciVis Contest\cite{scicon2015}. This particle data set is composed of 100 time steps with around 2,000,000 3D particles in each time step. Every particle has a position vector and seven-dimensional physical attributes: velocity vector, acceleration vector, and gravitational potential energy $phi$.
Along with the raw particle information, the data set comes with a list of \textit{halos}, which are features mainly defined by the particle density. The halos are found and tracked by the method proposed by Behroozi et al. \cite{rockstar}. We use the halo list to verify our feature extraction and tracking results based on the latent vectors.

\textbf{SNSPH}, SuperNova Smooth Particle Hydrodynamics, is our third particle dataset used to model core-collapse supernovae\cite{Fryer2006}. The dataset comes in a time series of 61 time steps and each time step contains around 900,000 particles. Features of interest are the finger-like structure reaching from the inner core to the outer core and the surface where these finger-like structures form. According to the domain scientists, these features are related to the temperatures and the densities of particles.

\subsubsection{Hyper-parameters}
\label{hyper_result}
\begin{table}[tb]
  \caption{
  Hyper-parameters used for different datasets. The particle positions in all datasets are normalized to be in the range $[0,1]$ and the radius in the table is the normalized radius. The number of samples shows the average sample size from each time step of the dataset. Attr. and Conc. stand for Attribute and Concentration respectively in the table.
  }
  \label{tab:hyper_comparison}
  \scriptsize
	\centering
  \begin{tabularx}{\columnwidth}{ccccccc}
  \toprule
    Dataset & Radius & Attr. & LatDim & \#Samples & \#Epochs & PSNR   \\
  \midrule 
	\multirow{4}{*}{Salt-LowRes}     & 0.05 & Conc. & 16 & 1,895 & 800 & \textbf{33.21}  \\
	                                & 0.01 & Conc. & 16 & 1,895 & 800 & 31.88  \\
	                                & 0.03 & Conc. & 16 & 1,895 & 800 & 30.21  \\
	                                & 0.08 & Conc. & 16 & 1,895 & 800 & 32.48  \\
  \midrule
	Salt-HighRes                & 0.008 & Conc. & 16 &16,175 & 500 & 32.65 \\
  \midrule
	DarkSky                     & 0.010 & $\phi ,v ,a$ &32& 20,529 & 900 & 35.48\\
  \midrule
	\multirow{3}{*}{SNSPH}    & 0.005 & $ T$ &20& 9,456 & 400 & \textbf{46.40} \\
	                            & 0.001 & $T$&16& 9,456 & 400 & 45.51\\
	                            & 0.030 &  $T$ &32& 9,456 & 400 & 44.30\\
  \bottomrule
  \end{tabularx}
\end{table}

To train the autoencoder on three datasets, we chose the hyper-parameters using the method discussed in \autoref{hyper_parameter_method} and sampled the training data using the method in \autoref{traning_testing}. To capture the feature variance in all time steps, we sampled around 1\% of particles in each time step for training. 
We treated the coverage of all samples in one time step as one epoch.
The hyper-parameters and training statistics for all datasets are presented in \autoref{tab:hyper_comparison}.

To validate the estimation method for determining the optimal patch radius, we trained four different models for the Salt Dissolution dataset and three different models for the SNSPH dataset. We choose the optimal radius and several smaller or larger radii to train these different models. Details for the radius chosen can be found in \autoref{tab:hyper_comparison}. It is worth noting that training multiple models is not necessary for our approach to determine the radius, we only did that to validate the LSCV estimation result. To qualitatively validate the estimated radius, we compare the data reconstruction and feature extraction between these models. In \autoref{tab:hyper_comparison}, we show that for both datasets, the optimal radius model resulted in better reconstruction peak signal-to-noise ratio (PSNR) than the model trained with a smaller or larger radius. This optimal radius also results in better feature extraction result as shown in \autoref{fig:main_results} (a).

Another noteworthy hyper-parameter is the latent vector dimensionality. The latent dimensionality shown in \autoref{tab:hyper_comparison} is determined by a random search technique. We found that for all the datasets, the optimal latent dimensionality is related to the number of particles in the particle patch and the attribute dimensionality we used in the experiment. Moreover, higher dimensionality will not decrease the model reconstruction quality, while slightly smaller dimensionality will only decrease the reconstruction moderately and have negligible influence on the feature exploration results.

\subsubsection{Reconstruction Quality}
\label{sect:recon}
\begin{figure}
 \centering 
 \includegraphics[width=\columnwidth]{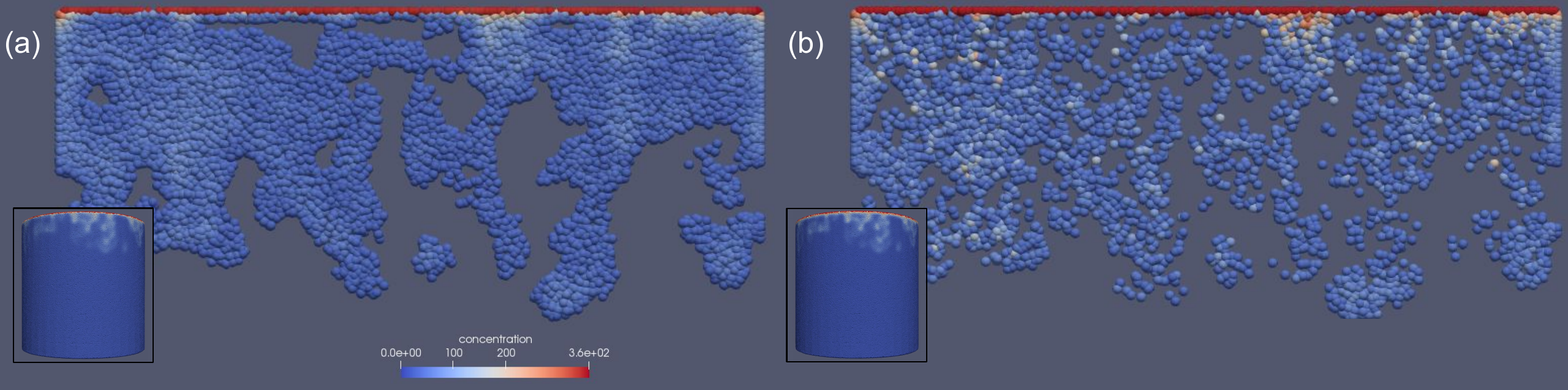} 
 \caption{
    Reconstructed (a) and original (b). Smaller images in the corner show the full data in one time step. The larger images show the particles filtered with the same salt concentration threshold. In (a) and (b), the color of the particle denotes its concentration value. 
 }
 \label{fig:reconstruction}
\end{figure}

We evaluate the reconstruction quality on the Salt Dissolution dataset.
In \autoref{fig:reconstruction}, we compare a zoomed-in slice along the z-axis of time step 35 in the low-resolution version. This slice was filtered to show only the particles with concentration values higher than 25. We can observe that the reconstructed concentration values are smoother among the reconstructed particles. It was found in the study\cite{scicon2016} on this dataset that these high-frequency concentration fluctuations are noise and are to be removed before feature extraction. Therefore, this smoothing effect in the reconstruction is beneficial to our subsequent feature extraction.
PSNR between the reconstructed and original data can be found in \autoref{tab:hyper_comparison}. Considering the beneficial smoothing effect, and that we only rely on the clusters of latent vectors to extract features but not to replace the original data with the reconstructed one,  PSNR values in this range are acceptable.

\subsection{Feature Exploration}

Feature exploration is the major use case for our latent vectors produced by the GeoConv autoencoder. The challenges faced in the three datasets are quite different. In the Salt Dissolution dataset, finger features are defined by the gradients in a scalar field. In the DarkSky simulation, halos are defined by the spatial distribution of particles. In the SNSPH simulation, features of interest are defined by a combination of spatial distribution and physical attributes. Through the experiments of these datasets, we demonstrate that the autoencoder-generated latent vectors can capture the salient features dictated by the applications.  

\subsubsection{Dataset 1: Salt Dissolution}

Since the viscous fingers are only related to the particle salt concentration value, we only include the concentration and particle positions to train the autoencoder. 
To begin feature exploration, we first choose time step 25 as our feature extraction example and the starting point of the tracking, where according to the prior studies\cite{salt_data} on this dataset, the number of viscous fingers starts to increase and their structure start to become complex.

\begin{figure}
 \centering 
 \includegraphics[width=\columnwidth]{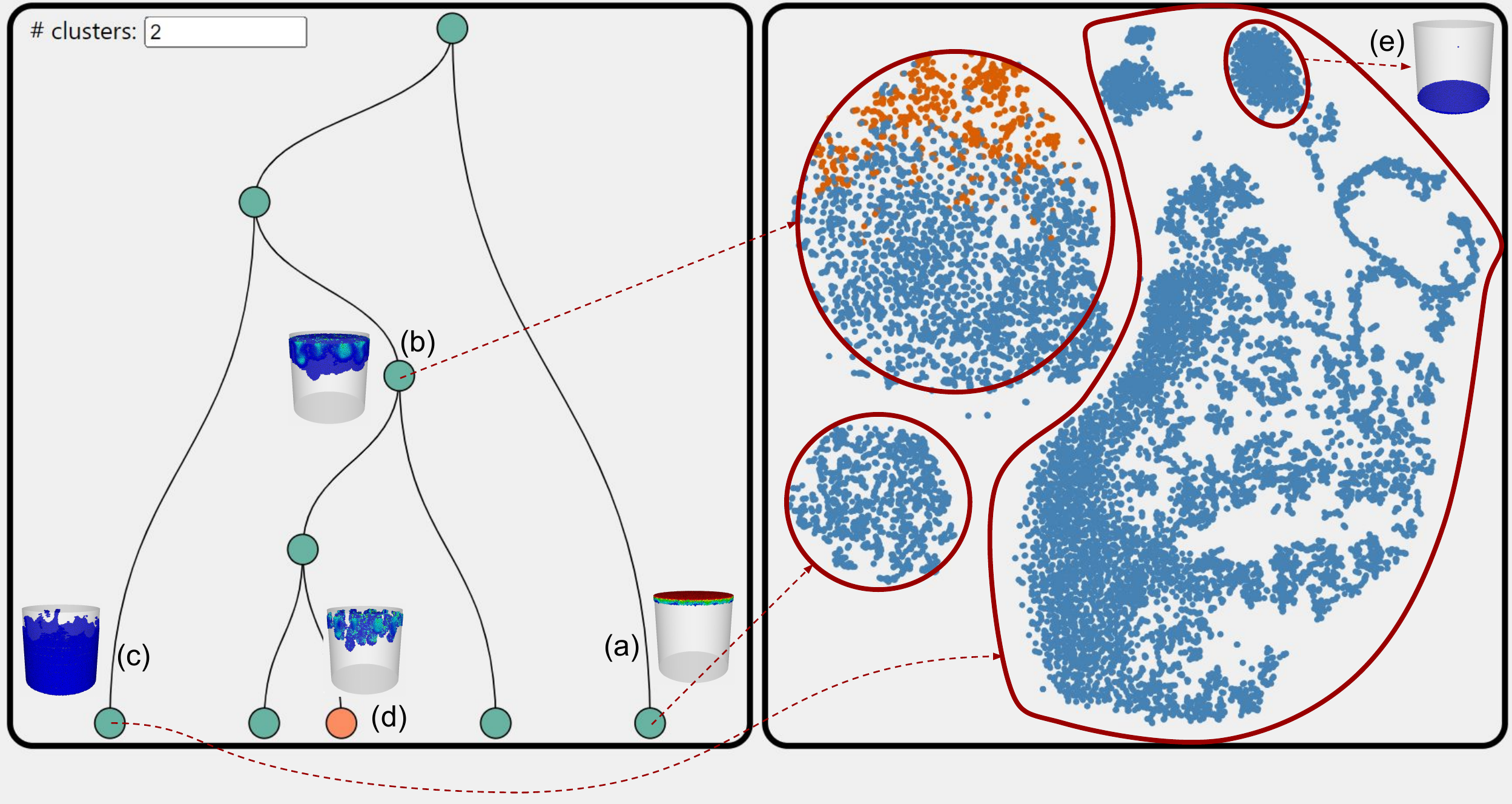} 
 \caption{
 The left panel contains the hierarchical clusters as the result of latent space exploration. The right panel shows the t-SNE projected particles. The selection of cluster nodes on the left panel will highlight the corresponding particles in the t-SNE view and visualize these particles in the original physical space, where the particles are colored by the salt concentration. 
 Features depicted are salt injection on the border (a), viscous fingers (b), background particles (c), thin cores inside of fingers (d), and a boundary not of interest (e)
 }
 \label{fig:fpm_vis}
\end{figure}

After we produce the latent representations, these vectors are loaded to our exploration system for further analysis. We present our feature exploration process in \autoref{fig:fpm_vis}. 
At the top level of hierarchical clustering, the dataset is clustered into the salt injection plane (\autoref{fig:fpm_vis} a) and the water body. We further split the water body cluster to find the finger structure, low concentration background, and the thin finger cores inside as shown in \autoref{fig:fpm_vis} (b), (c) and (d), respectively.
Circles and arrows in \autoref{fig:fpm_vis} show the correspondence between the cluster particles and their latent space 2D projection. This provides information on how distinct these clusters in the hierarchical cluster view are. However, these particle groups in the 2D projection can be misleading or may not necessarily be features of interest. For example, the small particle groups at the top of \autoref{fig:fpm_vis} (e) are corresponding to the boundary particles at the bottom of the water body and are not of interest. This makes it essential to combine both the latent space 2D projection and the physical space particle visualization to explore the features of interest interactively.

\begin{figure}
 \centering 
 \includegraphics[width=\columnwidth]{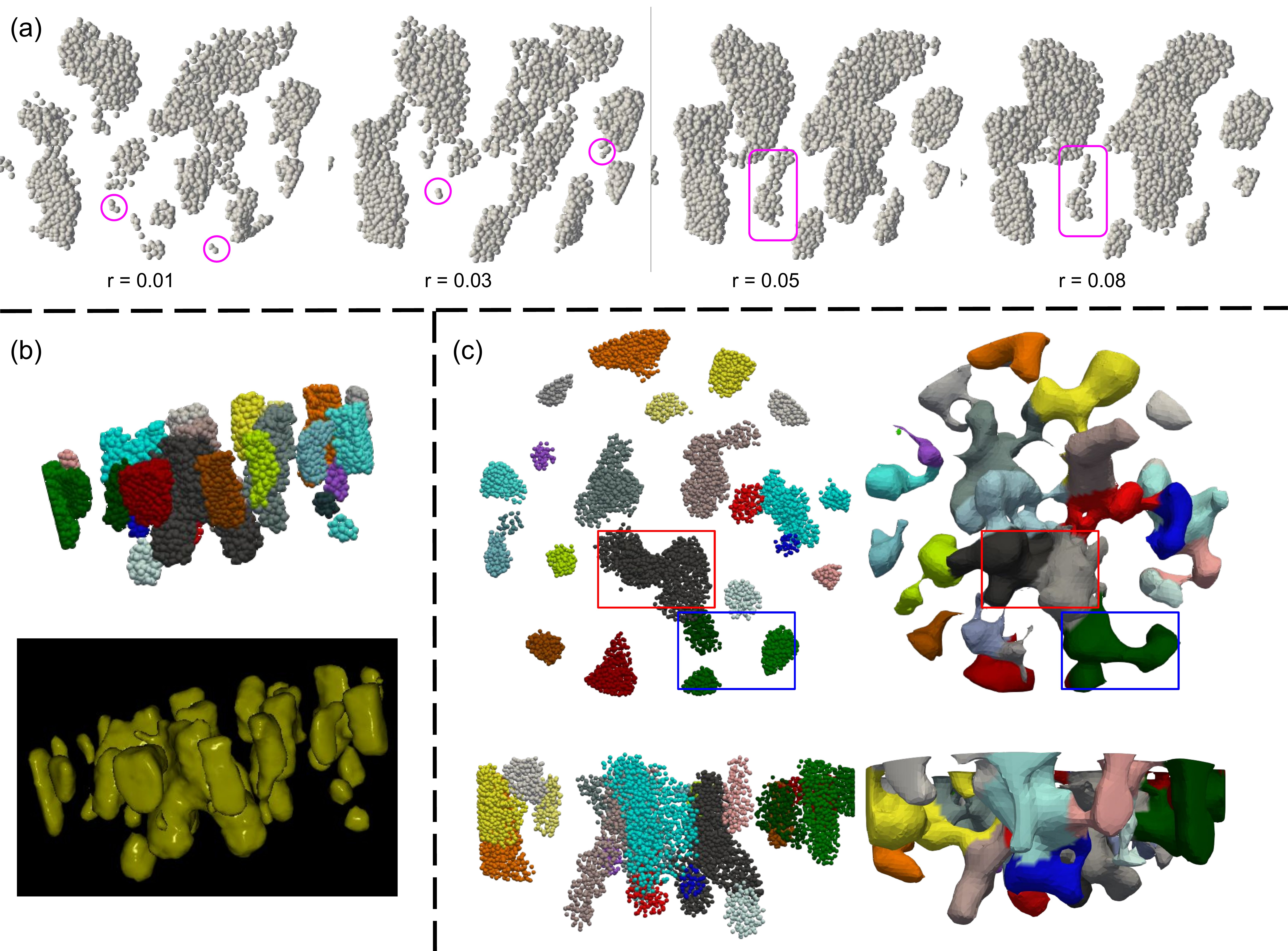}
 \caption{ 
 (a) Extraction results using four different models with different sizes of neighborhood considered. Notice noise (circles) when r=0.01,0.03, and the finger shrinking effect when r=0.88(squares).
 (b) Extracted fingers shown as particles and surfaces.
 (c) Different views for comparison between finger extraction through latent vectors and topology analysis. Fingers are colored to best match the same fingers from the two methods. It is observed two methods have different separation of fingers. Some fingers are grouped into one in our methods but divided using topology analysis (red square). Some fingers are in the reversed situation(blue square).  
 }
 \label{fig:main_results}
\end{figure}

We choose the finger core feature in \autoref{fig:fpm_vis} (d) for further feature analysis. 
\autoref{fig:main_results} (a) shows the finger cores both by direct particle rendering and surface extraction. 
In the direct particle rendering figure (top), we separated the finger cores with the Density-Based Spatial Clustering of Applications with Noise (DBSCAN), so that we can count the number of fingers and verify it both qualitatively and quantitatively by comparing to the features extracted by the winner of the 2016 SciVis contest. 
Surface extraction (bottom) is done by processing the particle patches defined on a regular grid. We assign 1 to the patches of finger cores and -1 to the patches of other clusters, and then extract the iso-surfaces at value 0 using marching cubes. 
The winners of the 2016 SciVis contest proposed two approaches to extract the finger structures \cite{scicon2016}. The first one is based on topology analysis on the data and the second one uses slice-wise clustering. We ran the topology-based method and our method on the same data of time step 25. The finger extraction results are shown in \autoref{fig:main_results} (b). The two methods generate a similar numbers of fingers; 24 are generated from the topology analysis, and 25 are generated from our method. We colored the fingers to match the extracted results using the two methods. However, there are still some noticeable differences between the two results.
One observation from the result is that topology analysis generates fingers of larger size. This is likely because the analysis is based on a smooth distance field, which will connect the boundaries between fingers. 
Secondly, topology analysis and our method separate the fingers differently in space. Since we use DBSCAN in our method to separate them, connected fingers with the same particle density will be identified as one finger. This effect can be seen in \autoref{fig:main_results} (b) within the red squares. On the other hand, the topology analysis method only uses persistence thresholds to find fingers, which will fail to detect some of them and merge the corresponding region into a single finger as shown in the blue squares. 
Overall, the finger extraction quality using our proposed approach is comparable to the feature extraction result by topology analysis.

\begin{figure}
 \centering 
 \includegraphics[width=\columnwidth]{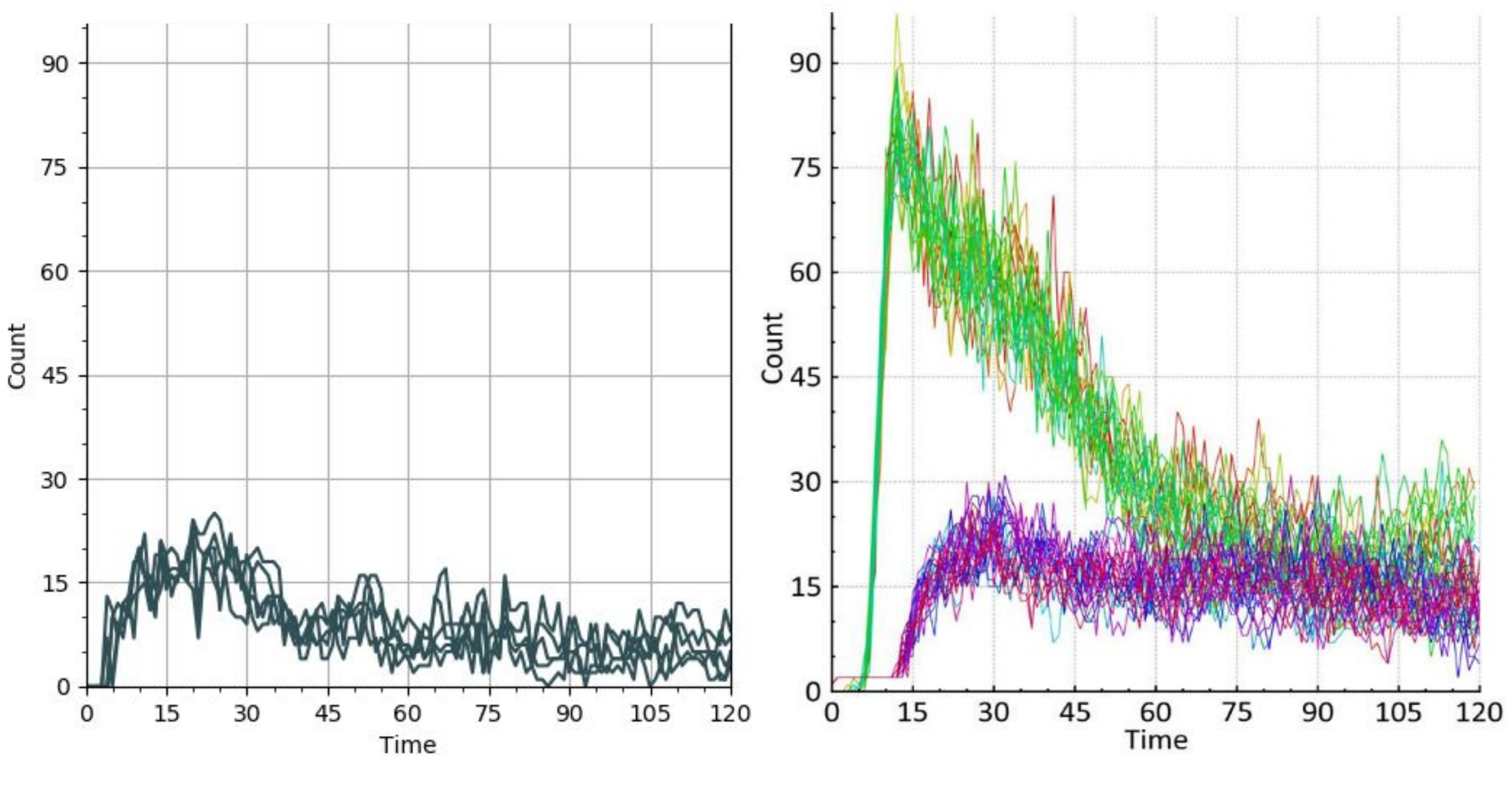}
 \caption{
Left: The extracted finger count across all 120 time steps on 5 different ensemble runs using our method. Right: Finger count from the 2016 SciVis contest winner \cite{scicon2016}. Blue and purple lines are from the heuristic clustering method in their work, while green and yellow lines are from topology analysis.
 }
 \label{fig:compare_topology}
\end{figure}

By comparing with the methods\cite{scicon2016} specifically designed for this dataset, we verify our proposed approach can capture the target features defined by the particle physical attributes successfully. 
Across all time steps, our method has a similar trend in finger count to both methods we compare with, and we have a very similar finger count to the heuristic clustering result as shown in \autoref{fig:compare_topology}. Detected finger count both go up after the first 30 time steps and then drop when the data become chaotic in the latter time steps. 
It was explained in their work that the number difference between the topology analysis and heuristic clustering is mainly because of the choice of threshold. 

\subsubsection{Dataset 2: Dark Sky Simulation} 

\begin{figure}
 \centering 
 \includegraphics[width=\columnwidth]{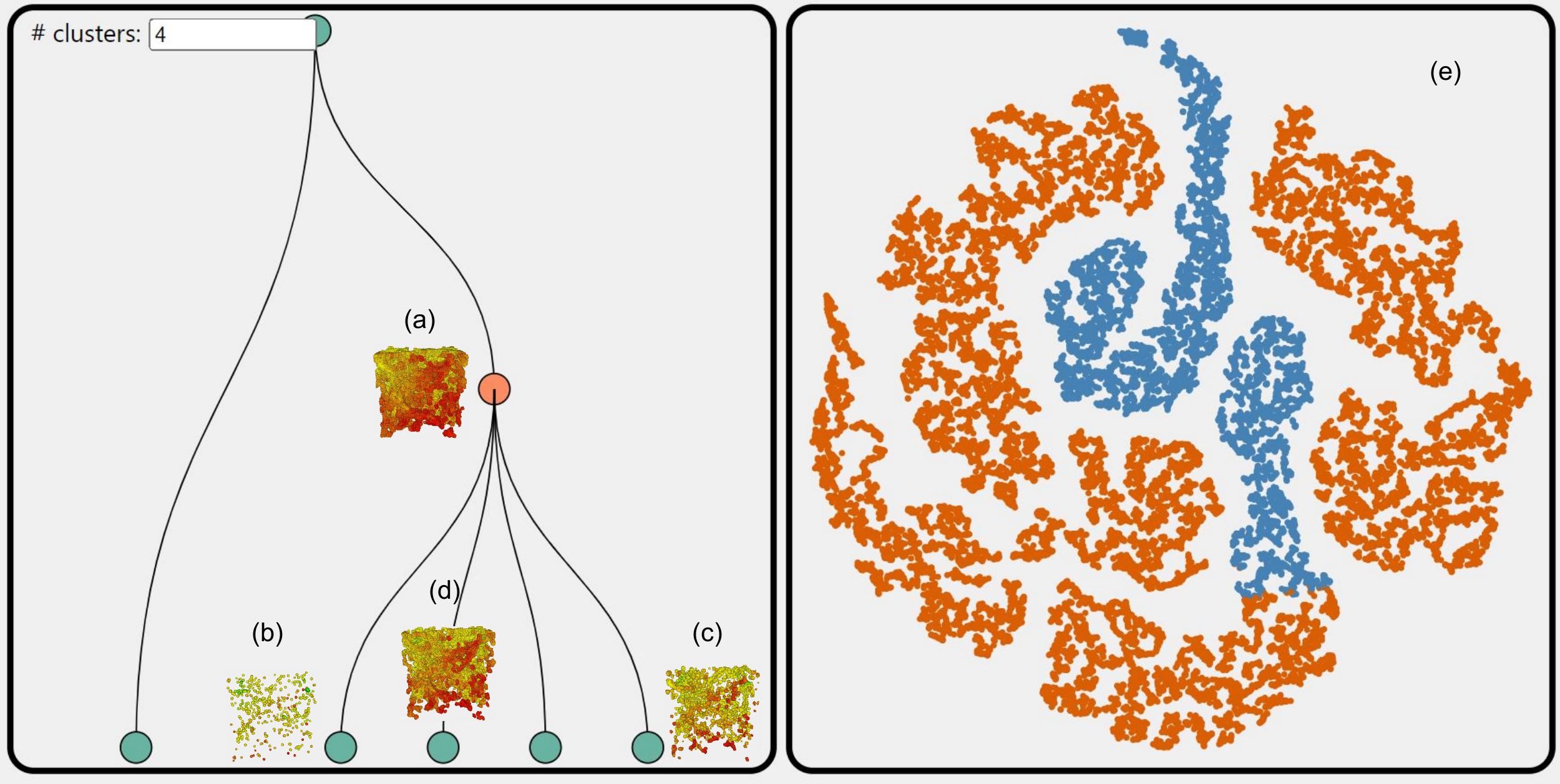}
 \caption{ 
 Feature exploration for time step 49. The figure on the right shows that cluster (a) is corresponding to about four (or more, based on how the user perceives the group connectivity) particle groups in the latent 2D projection. We subdivide the cluster (a) into four lower-level clusters and found that three of them (b), (c), and (d) may contain the halo.
 }
 \label{fig:cos_vis}
\end{figure}

\begin{figure}
 \centering 
 \includegraphics[width=\columnwidth]{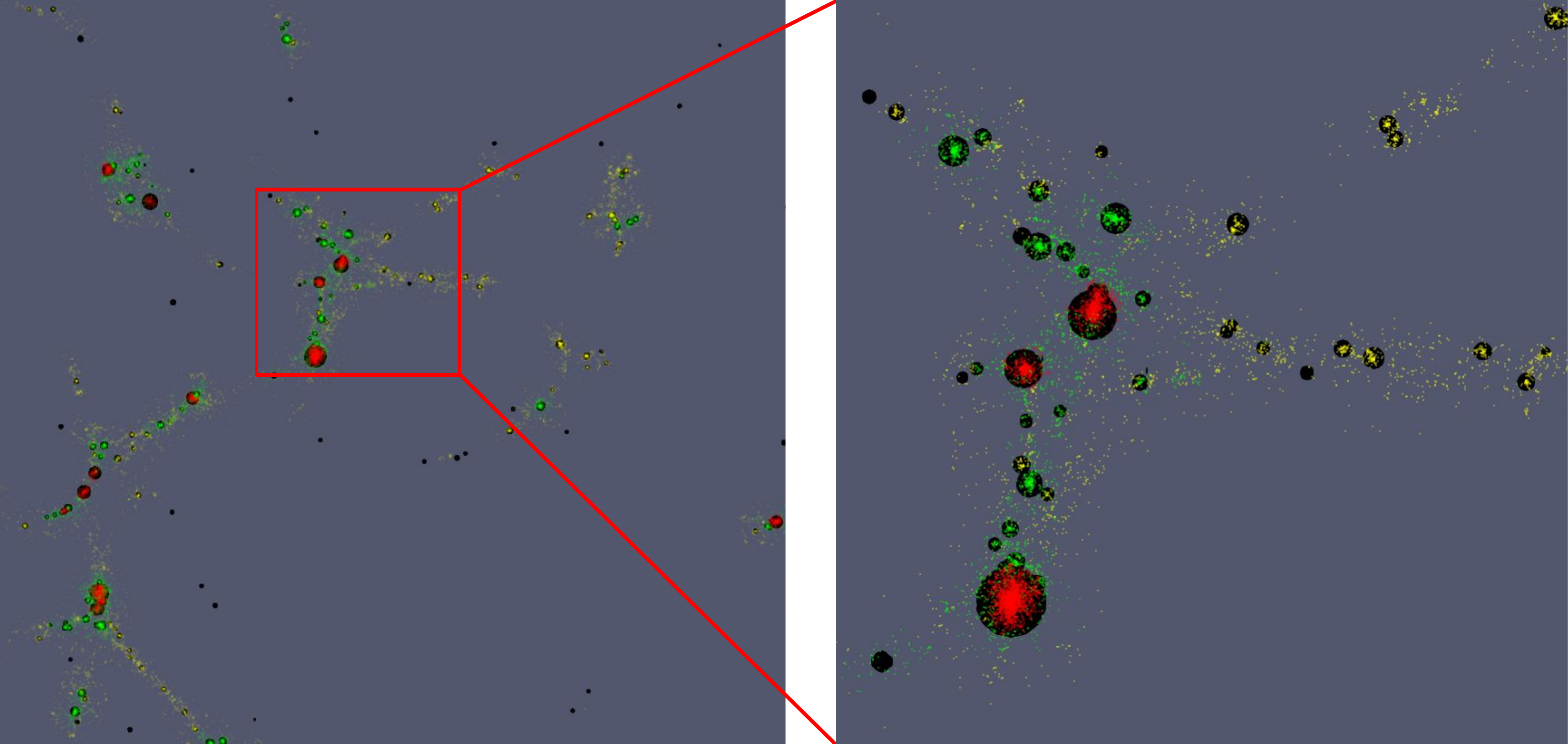}
 \caption{ 
 One slice of the data from time step 49. Particles in \autoref{fig:cos_vis} (b), (c) and (d) are colored with red, green and yellow in this figure. We overlay the ground truth halo position and size with black spheres. Large halos are captured by the red cluster and smaller halos are mostly in the green and yellow clusters.
 }
 \label{fig:cos_combine}
\end{figure}
Algorithms for locating halos, which use hand-crafted feature descriptors for dark matter halos, primarily focus on the distribution of the particle locations to find the regions of high particle density. The velocity of particles is used in some of the algorithms to exclude the particles with a velocity that exceeds the local escape velocity. Therefore, we include the particle position and velocity to train the autoencoder. Moreover, to make the feature extraction task more challenging, and to verify that our approach could work in situations when the feature-related physical attributes are not known, we also include the acceleration and gravitational potential energy $phi$ in the training. 
After the autoencoder is trained, we followed the same process described above to find the interesting features. In the first level of hierarchical clustering, we found the target feature may be contained in node shown in \autoref{fig:cos_vis} (a). There are approximately four (or more, based on how the user perceives the group connectivity) distinct particle groups in the 2D projection of the latent vectors in cluster (a). Therefore, we subdivide the cluster (a) into four lower-level clusters. Based on the physical space visualization of these four clusters, we found three of them contain the target feature of halos. 
In the case the user chooses a cluster number of more than four and child nodes do not depict desired features, the clustering and splitting step can be easily revoked and the user can change the number of clusters. 

To verify the overlap of identified clusters and the ground truth halo features, \autoref{fig:cos_combine} shows each particle colored based on their cluster, while the real halos' position and size are drawn as black spheres over the particles. The red cluster (\autoref{fig:cos_vis} b) defines the particles in the large halos and the other two clusters mostly include the smaller halos. However, A close look at the left figure in \autoref{fig:cos_combine} shows that there are still some halos with a very small size which are not captured by our method.
We calculate the percentage of halo particles that are covered by the chosen clusters, where higher percent is better. The red and green clusters along cover 74.9\% of the halo particles across all time steps, while all three clusters (yellow, green, and red) cover 91.9 \%. Even though the clusters found by the exploration tool cannot perfectly extract all the halos, the latent representation at least correctly provides the decision boundaries related to the particle density, especially when the irrelevant attributes are included in the purpose to make the task harder.

\subsubsection{Dataset 3: SNSPH}
With the final dataset, our goal is to verify that our method can capture the features defined by both the particle spatial distribution and physical attributes. 
As mentioned in the dataset description, SNSPH models a core-collapse supernova. In the process of supernova formation, the inner part of the stellar core is compressed into neutrons. This causes the later infalling material to bounce and form an out-racing shock wave. Scientists are interested in the viscous fingers formed along with the shock wave propagation. \autoref{fig:supernova} (a) shows a ring of high density. The inner core is roughly enclosed by this ring, and the outer core refers to other parts in the figure.
Viscous fingers in this dataset are defined both by temperature and particle spatial distribution (density and the spherical shape). Thus, although there are a total of 91 physical attributes in this dataset, we only choose the temperature to train the neural network.
\begin{figure}
 \centering 
 \includegraphics[width=\columnwidth]{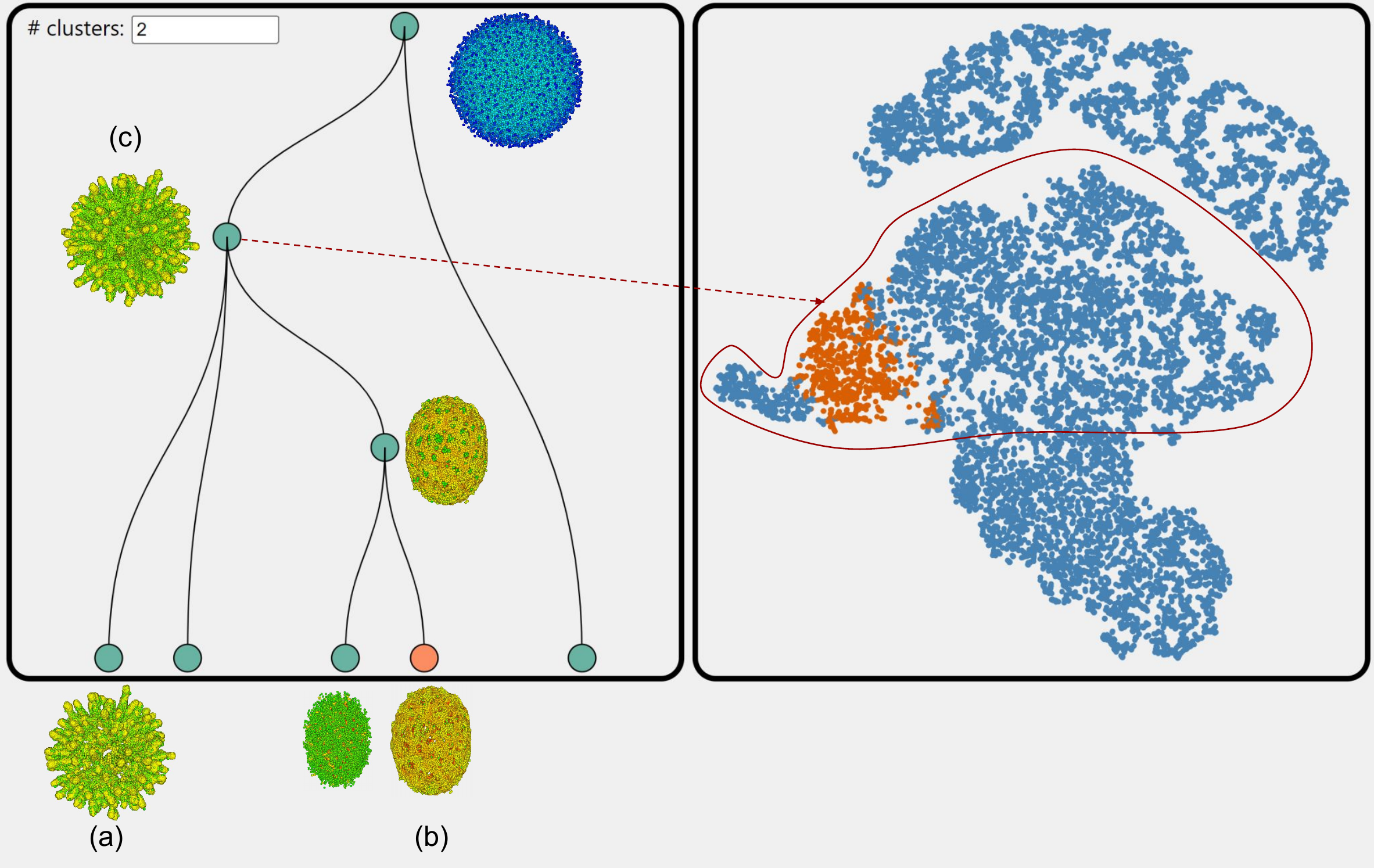}
 \caption{
 Left: Hierarchical Cluster View. Right: Latent Space Projection View. Particle visualization in the physical space is only shown in the pop-up windows when the cluster node is selected in the exploration process. Here, particle visualizations are placed beside the cluster node in the left figure for comparison. The latent space projection view also highlights the particles from the selected cluster node.
 }
 \label{fig:jet3b_explore}
\end{figure}

\begin{figure}
 \centering 
 \includegraphics[width=\columnwidth]{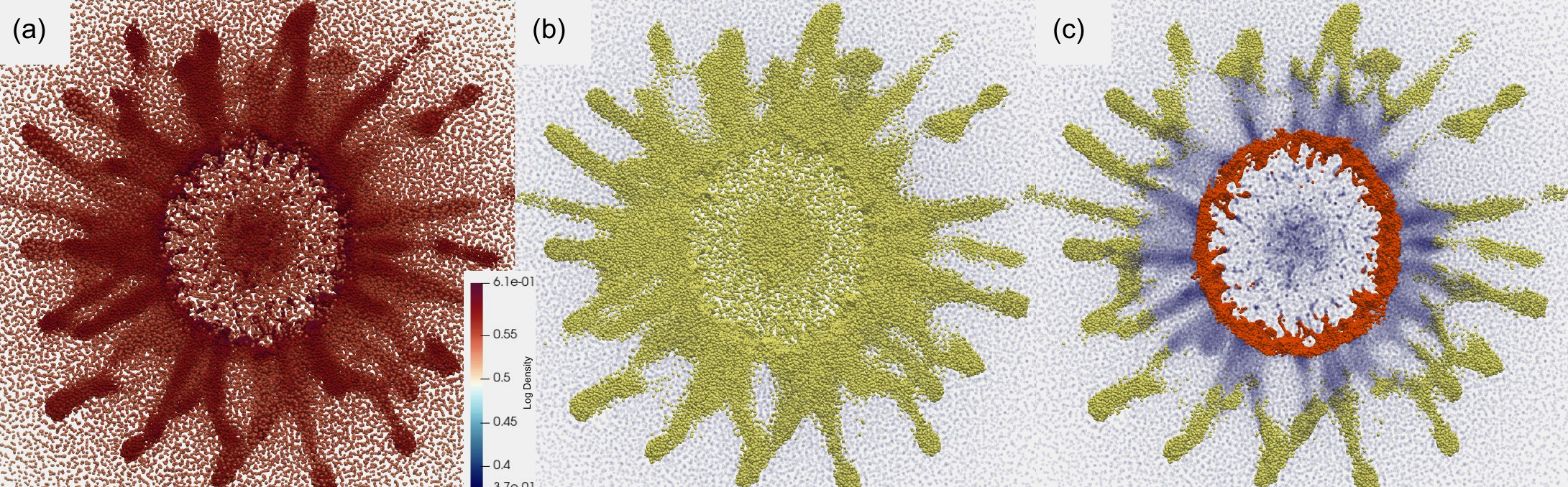}
 \caption{
 (a) 2D slice of the supernova simulation with particles colored based on the density. 
 (b) Slice with highlighted particles corresponding to \autoref{fig:jet3b_explore} (c).
 (c) Slice with highlighted particles. Orange particles correspond to \autoref{fig:jet3b_explore} (a) while yellow particles correspond to \autoref{fig:jet3b_explore} (b).
 }
 \label{fig:supernova}
\end{figure}
Using the proposed latent space exploration system, we successfully found two interesting features in this dataset. Our exploration result is presented in \autoref{fig:jet3b_explore}. The first split of the hierarchical tree already revealed some interesting structures in the dataset. Through the physical space visualization, we find the left-side node cluster shows the finger-like structure (see \autoref{fig:supernova} (b) for a close-up slice visualization). However, this cluster also includes the particles inside the inner core.
Since the domain scientists are interested in the more detailed structure inside the fingers and inner core, we further split the left-side cluster into the second level of hierarchical clustering. Repeating these steps, we successfully find the clusters corresponding to the features of ``out-reaching finger'' structures as shown in \autoref{fig:jet3b_explore} (a)  and the surface between the inner and outer core, where these fingers start to form as shown in \autoref{fig:jet3b_explore} (b).
A close up physical space visualization is shown in \autoref{fig:supernova} (c), where we highlight the ``out-reaching finger'' feature (\autoref{fig:jet3b_explore} a) in yellow and the finger formation surface (\autoref{fig:jet3b_explore} b) in orange.
Since these two features are on different levels of the hierarchical clusters, it is hard to detect both without a hierarchical division of the clusters.

Though the particles forming these two features have high particle density in their regions, the feature depicted in \autoref{fig:jet3b_explore} (b) is defined additionally by high temperature near the supernova inner core and the spherical shape between the inner and outer core. Without the latent vectors produced by the autoencoder in our approach, it is tedious to manually search for the decision boundary in temperature and density to separate these features.

\subsection{Comparison with Other Particle Patch Representation}

\begin{figure}
 \centering 
 \includegraphics[width=\columnwidth]{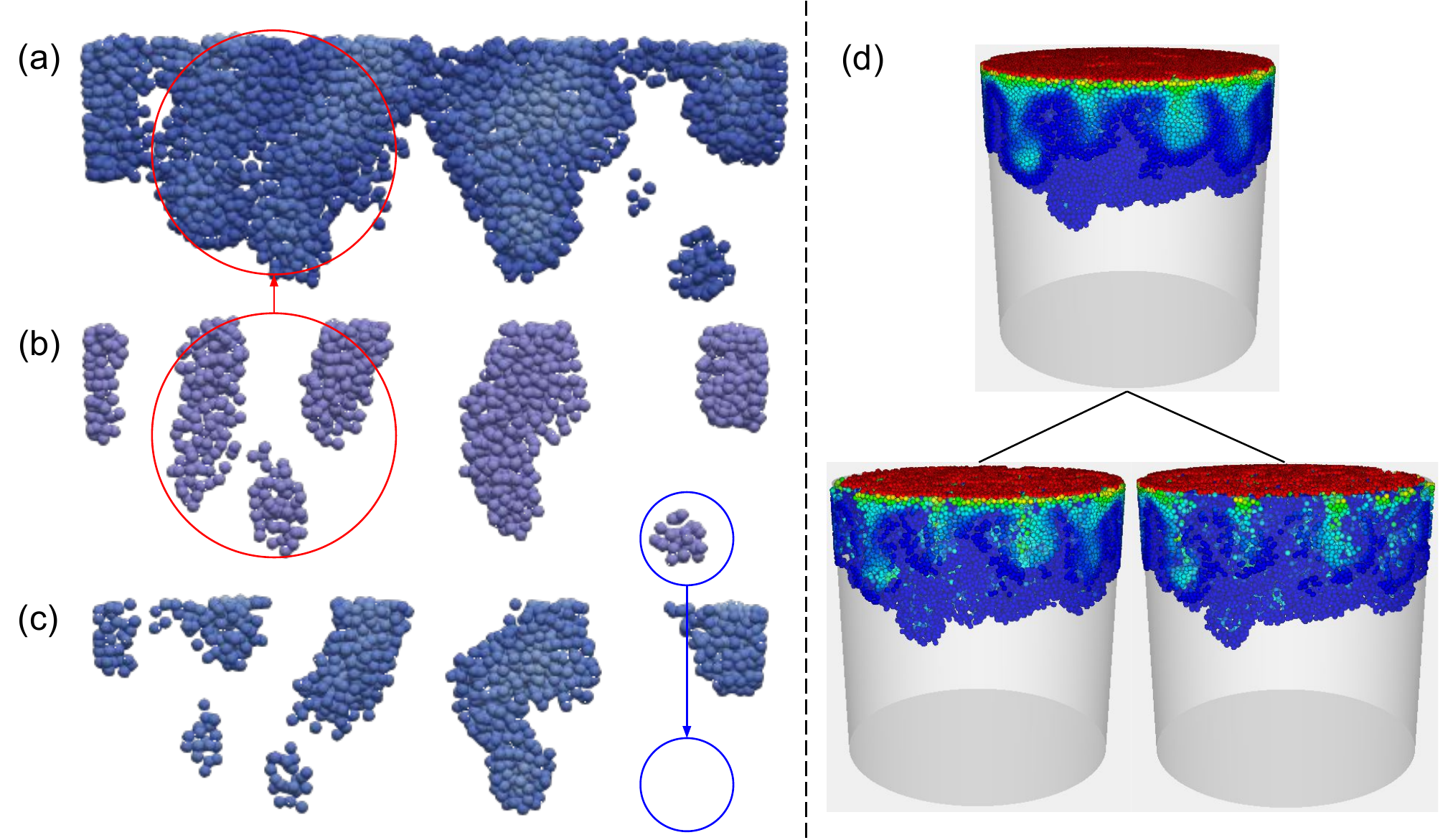}
 \caption{
 (a) Average the particle's concentration with particles in a local neighborhood and set a low concentration threshold to extract features. (b) finger core extraction result with latent vectors. (c) The same average set as (a), however, set a high concentration threshold. (d) Three cluster nodes from hierarchical clustering result of PCA. The top cluster shows the high concentration particle clusters found by PCA representation, however, we cannot further divide this cluster to find the finger cores.
 }
 \label{fig:compare_mean_pca}
\end{figure}

Autoencoders are not the only way to represent the particle patches. In this section, we are going to show that latent vectors produced by the autoencoder are more informative compared with two other methods: representing the patch with a mean in the neighborhood and with principal components. We use the salt dissolution dataset to perform the experiments. 
Some prior knowledge about the viscous fingers is that they only appear in the region that has relatively high salt concentration. A straightforward way to extract fingers is to filter the particles with a concentration threshold.
Therefore, the first method we compare is averaging the concentration in the neighborhood to remove the high-frequency noise and extract features with a threshold.
The comparison result is shown in \autoref{fig:compare_mean_pca} (a), (b) and (c). It demonstrates the difficulty to determine the threshold if we only remove the noise with an averaged concentration in the particle patches. A low threshold will make it difficult to distinguish different fingers, while a high threshold will make some fingers disappear. Our neural network method has the advantage of capturing features defined by the change of concentration automatically. It is probably true that we can identify features by calculating concentration gradients and there have been previous work\cite{Xu2019} using gradients to identify viscous fingers in a volumetric dataset. However for particle data, calculating gradients is difficult and can be expensive to apply mesh-free methods to achieve this on particles.

Another way is to perform principal component analysis (PCA) on the particle patches and represent each patch with its principal components. 
We treat each particle as a data sample with four-dimensional variables (three dimensions of position and one dimension of concentration). So we use the four principle components as feature descriptors for the patch. Four principal components are ordered by their explained variance and concatenated into a vector.
In \autoref{fig:compare_mean_pca} (d), we can find the high concentration clusters which contain our target feature fingers with our hierarchical clustering exploration process.
However, further division of this cluster did not reveal any finger cores inside the high concentration region. 
Particles inside the high concentration region seem to be only randomly assigned into two child clusters. This indicates that principle components of the particle patch failed to capture the information related to the gradient of concentration, which is crucial to define the finger cores and is captured using our method.

\begin{figure*}
 \centering 
 \includegraphics[width=\textwidth]{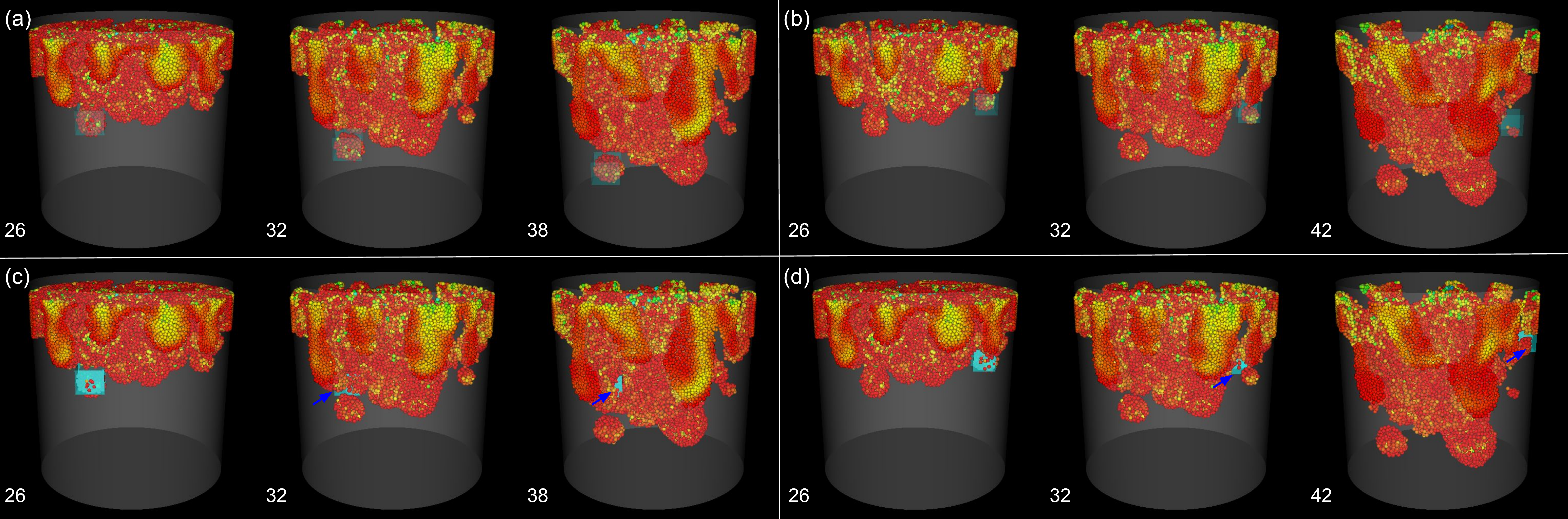}
 \caption{
Time snapshots from two ``finger tip" tracking example through time step 25 to 45. (a) and (b) use the latent vectors generated with our approach. (c) and (d) use the raw particle attributes. Time is labeled at the bottom left corner of the image. The user selection of the feature of interest is highlighted by the blue square in the figure.  Particles are colored by their concentration values. }
 \label{fig:tracking}
\end{figure*}

\subsection{Feature Tracking}

As another use case for our autoencoder-generated particle patch representation, successful tracking of features with the mean-shift algorithm in the latent space is demonstrated in this section. In the salt dissolution dataset, since we do not have the ground truth tracking results, we verify the tracking qualitatively by showing the features and the tracking bounding box. As for the DarkSky dataset, we directly compare the tracking result with the halo merger tree generated by \cite{takle2012tracking}. We also compared the latent vector based mean-shift tracking with the original physical attributes based mean-shift tracking in these two datasets.

\subsubsection{Viscous Finger Tracking}
We select two fingertips from time step 25 and track them consecutively for 20 time steps. In \autoref{fig:tracking} (a), we present the tracking box at time step 26, 32 and 38 using our latent vectors. The tracking bounding box follows the ``fingertip" feature with only a small error.
As comparison, we show the tracking bounding box at the same three time steps 26, 32, and 38 using the raw physical attributes in \autoref{fig:tracking} (c).
In the latter two time steps, the tracking box completely went off the feature and was partially blocked by the other particles (indicated by the arrow).
In the other fingertip tracking example with latent vectors (\autoref{fig:tracking} b) and raw attributes (\autoref{fig:tracking} d), the feature shrinks as time advances, and eventually disappears at time step 45.
Compared to the tracking with raw physical attributes, our method still provides better tracking stability. 
We also provide the tracking animations for these two examples in the supplementary materials. 

Generally speaking, tracking when features are small or time resolution is low is more challenging because the mean-shift tracking algorithm we choose relies on the overlap of features between time steps. However, the tracking results prove that our generated latent vectors capture the important features which allow for more accurate and stable feature tracking. The possible reason why the latent vectors work better than the raw attributes in tracking is that the latent vectors encode the features in a more compact space, where some unstable attributes or noise are smoothed or neglected. This effect also helps us easily separate the features in the latent space.

\begin{figure}
 \centering 
 \includegraphics[width=\columnwidth]{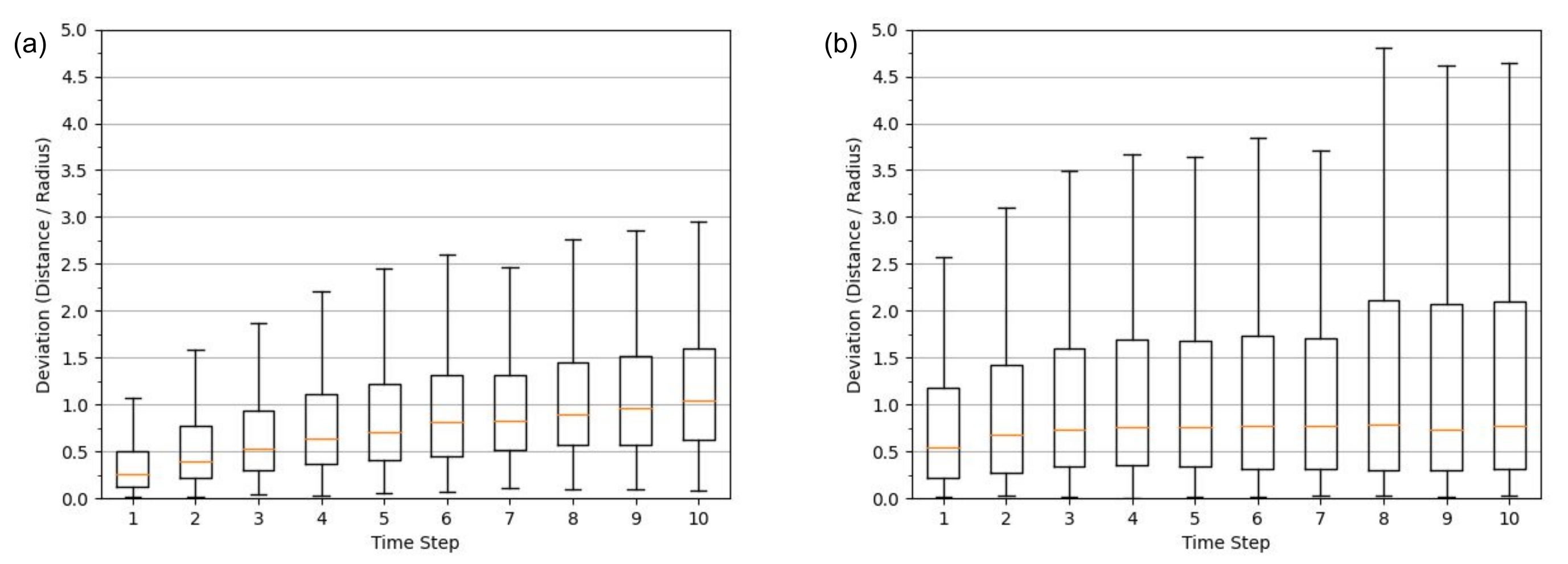}
 \caption{
  Tracking deviation for ten time steps from 594 random halos shown with a boxplot. (a) Tracking with latent vectors generated using our method. (b) Tracking with raw attributes.
 }
 \label{fig:deviation}
\end{figure}
\subsubsection{Halo Tracking}
For this dataset, Since we had the features' ground truth positions, it is easy to evaluate our feature tracking results by calculating the distance from the center of the extracted feature produced by our method and the real feature center. Similar to the tracking experiment in dataset 1, we also compare the latent vector tracking against using the raw attribute for tracking.
Taking the feature size into account, we use $\frac{d}{r}$ as the feature deviation metric, where $d$ is the distance between the centers and $r$ is the ground truth halo radius. If this metric is larger than 1 but smaller than 2, the center of our tracked halo is no longer within the ground truth halo, meaning a slight deviation. If larger than 2, the tracking center and the feature center will have no overlap.
We randomly chose 594 halos from the halo list and tracked them for ten time steps. The results of the tracking error are shown in \autoref{fig:deviation}. For both methods, the median deviation generally increases with time and reaches one after ten time steps. In the first five tracking time steps, tracking accuracy with latent vectors is significantly better, while in the last few time steps the tracking accuracy using raw attributes is marginally higher. However, the tracking stability using latent vectors is significantly better than using raw attributes, as the range between the first and the third quartile in latent vector tracking is much smaller.

\subsection{Computation Time and Memory Footprint }

In this section, we report the computation time and memory consumption for different parts of our proposed approach from the experiments on the three datasets. The machines we used for experiments are as follows: Our autoencoder model is trained on a supercomputer node with 2.4GHz Dual Intel Xeon 6148 (40 cores, but we only use one core in training) and NVIDIA V100 GPU with 16GB video memory. While the extraction and tracking experiments are done on another machine with an 8-core Ryzen 2700X, 32GB of RAM, and an NVIDIA RTX 2060 GPU.

We present the total training time, latent vector inference time, clustering time, t-SNE projection time, and tracking time in \autoref{tab:timing}. The model training time is highly related to how many particles we sample in each time step, particle patch size, and how many epochs we train before the network converges. These parameters can be found in \autoref{tab:hyper_comparison}. The autoencoder training needs to be done only once for each dataset. Due to the relatively small number of layers in our network and the training data sampling strategy, our approach's training time is shorter than other works which also use neural networks to extract features \cite{FlowNet,Cheng2019}. The inference time shown is for generating all particles' latent vectors in a single time step (i.e. forward pass of a trained encoder). This inference process is necessary for latent space exploration and is linear to the number of particles in one time step, which will scale well on larger datasets. 

The clustering time shown in the table is calculated by applying the k-means clustering on all the particle latent vectors in one time step, which is the same as the first level clustering in our hierarchical cluster process. This time provides an approximate upper bound for all the user-decided clustering time during the feature exploration process. This quick clustering time facilitates interactive exploratory visualization. Since the k-means time complexity is also linear to the number of particles, this clustering time will also scale well to larger datasets.
All particles' latent vectors in one time step are usually too large for t-SNE projection in a reasonable time. As discussed before, the t-SNE projection time shown is for around 1\% of all particles. The projection of 20,000 samples from one time step in the DarkSky dataset takes 66.5 seconds. This projection is a preprocessing step before feature exploration and will not influence the system's interactability. 
Since t-SNE projection does not scale well to a large number of instances. There are two possible solutions if we want to apply our approach to larger datasets: (1)  Sample the particles more aggressively to keep a smaller number of samples. (2) Adopt other more efficient projection algorithms such as UMAP\cite{umap}. 

We calculate the computation time for the topology analysis we use to compare our method against for dataset 1. The topology analysis takes 62.5s for the high-resolution version of the dataset.
Our approach's preprocessing time (inference + t-SNE projection) for this dataset is on the same order. From the original halo finder work \cite{rockstar}, we calculate that their processing time for one time step in the DarkSky simulation is approximately 72s. Considering their time is measured on a legacy 2009 machine (AMD Opteron 2376), the computation time of our method is longer than the original halo feature finder algorithm on this dataset. However, these comparisons only include the actual computation but not the exploration and engineering effort to design the suitable feature descriptor in the comparison methods. Our proposed approach can achieve this exploration semi-automatically at the cost of around 1 hour of model training and seconds of extra preprocessing time.

\begin{table*}[tb]
  \caption{Training, inference, clustering, tracking time, and size for the three datasets.}
  \label{tab:timing}
  \scriptsize
	\centering
  \begin{tabularx}{0.8\textwidth}{ccccccccc}
  \toprule
    Dataset	& \#Particle & Training & Inference & Cluster & t-SNE & Tracking & Network Size & Latent Size \\
  \midrule 
	Salt Dissolution-Low Resolution & 192k & 8.4 min & 14.3 s & 1.9 s & 5.4 s & 90 ms & 285 KB & 11.7 MB \\
	Salt Dissolution-High Resolution & 1,664k & 38.4 min & 59.1 s & 6.7 s & 50.8 s & - & 285 KB & 101 MB\\
	DarkSky & 2,097k & 1.25 h & 81.4 s & 8.5 s & 66.5 s & 107 ms & 302 KB & 228 MB \\
	SNSPH & 946k & 23.6 min & 38.2 s & 4.8 s & 44.1 s & - & 289 KB & 57 MB\\
  \bottomrule
  \end{tabularx}
\end{table*}

\begin{figure}
 \centering 
 \includegraphics[width=\columnwidth]{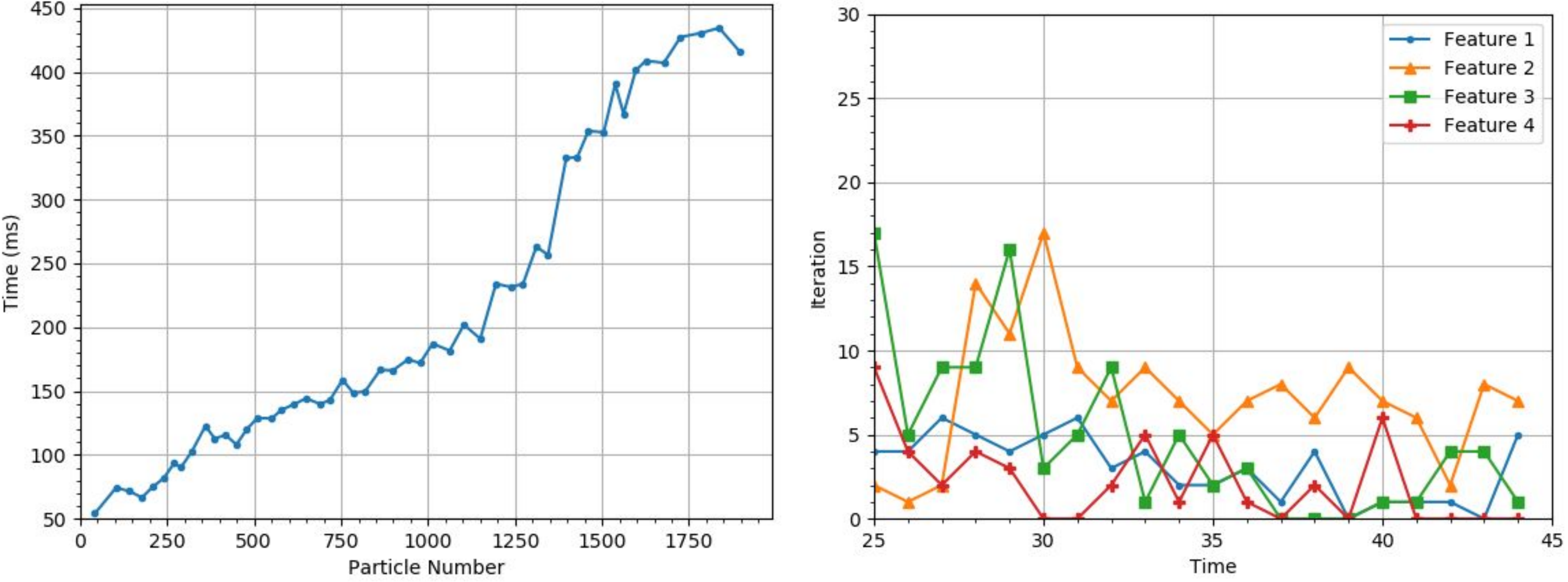}
 \caption{
 Left: Relations between the number of particles in the feature region and feature tracking time in the Salt Dissolution dataset.
Right: Number of iterations before convergence for 4 different feature tracking examples. Feature 1 and 2 are the same features as shown in \autoref{fig:tracking}.
 }
 \label{fig:iterations}
\end{figure}

Lastly, in the process of tracking, the number of iterations before the mean-shift algorithm converges is similar for different features. Therefore, the feature tracking time depends on the size of the feature. A larger feature size means more latent vectors need to be generated. The total tracking time is dominated by the latent vector generation time. For dataset 1, in the case that the feature of interest is a fingertip, which is consisted of approximately 200-400 particles, the tracking time is around 90ms. 
In \autoref{fig:iterations}, we present the results that show a linear relationship between tracking time and feature size, as well as the number of iterations before convergence.

As for the memory consumption, our method samples the whole dataset and processes the samples in a patch-by-patch manner. Because of this and the relatively small network size, our model can be trained on GPU with lower memory (for example, RTX 2060 with 6GB GPU memory) if we decrease the mini-batch size in training. 
We provide the size of networks and latent vectors for different datasets in \autoref{tab:timing}. When it comes to an even larger dataset where computation time and memory are issues, we can reduce the sample density in the latent vector generation stage to make the method scalable.

\section{Discussion, Limitation, and Future Work} 
\textbf{Usability}. One challenge for particle feature extraction is tedious feature descriptor design for the specific dataset. We overcome this challenge by automatically learning feature descriptors in a data-driven way. With the help of hyper-parameter estimation methods and an interactive hierarchical clustering system, our approach can be easily used by domain scientists. They only need to provide the knowledge on which physical attributes are related and interactively identify the features extracted by the autoencoder.

\textbf{Scalability}. In the experiments, our method works within reasonable computation time and memory consumption on the dataset containing up to 2,000,000 particles. However, the performance on larger datasets could be an issue. This scalability issue is mostly caused by the need to produce a latent vector for every particle in a time step before exploration. This issue could be alleviated by sampling a subset of particles to produce latent vectors, which may lead to reduced extraction quality or artifacts. 

\textbf{Interpretability}. 
Even though latent space visualization methods such as our approach and others \cite{FlowNet,v2v,porter2019deep,Cheng2019} are helpful in different applications, latent spaces are still not fully understood.
We provide some intuitions and evidence on the characteristics of our latent spaces, such as how latent representations remove noise and extract high-level features. How to understand neural network based latent vectors and why distances in the latent space are useful are still open questions.

In the future, scalability and interpretability are two fields where we want to explore. Compression methods that quantize the latent vectors in \cite{Cheng2019} could be a solution for the scalability issue on large datasets. Another possible way is to sample the representative particle patches in the datasets, which will reduce the computation cost. How to define representative patches and design a sampling technique are the keys to this problem.
Regarding latent space interpretability, studies in the field of explainable AI provides some insights. Visual analytics systems can be useful to probe the encoded high-level features in different dimensions of the latent space. Understanding the encoded features will help us diagnose and improve the model.

\section{Conclusion}
In this paper, we show that latent vectors produced from a GeoConv based autoencoder are useful for feature extraction and tracking.
We demonstrate successful feature extraction results in three different particle datasets and show how mean-shift tracking is useful to track features through the latent space for two of these datasets. 
In the first dataset, we show our latent vector can reconstruct particle patches with less noise than raw data. The extraction results for viscous fingers, which are defined by concentration, are superior to the threshold method or representing the neighborhood with principle components. We validate the extraction results by comparing them with the SciVis Contest winner's approach. 
In the second and third datasets, we show that the latent representations can capture the features related to particle spatial distribution and physical attributes.  
Finally, the detailed performance of our approach is provided. The efficiency of the approach is comparable to the benchmark methods, and the effort to design and engineer specific feature descriptors for the dataset is not required in our approach. 

%



\ifCLASSOPTIONcompsoc
  \section*{Acknowledgments}
\else
  \section*{Acknowledgment}
\fi

This work is supported in part by the National Science Foundation Division of Information and Intelligent Systems-1955764, the National Science Foundation Office of Advanced Cyberinfrastructure-2112606, U.S. Department of Energy Los Alamos National Laboratory contract 47145, and UT-Battelle LLC contract 4000159447 program manager Margaret Lentz.




\bibliographystyle{IEEEtran}
\bibliography{template} 
\end{document}